\documentclass{article}
\usepackage{amsmath,multirow,color}
\usepackage{natbib}
\usepackage{hyperref}       % hyperlinks
\usepackage{url}            % simple URL typesetting
\usepackage{booktabs}       % professional-quality tables
\usepackage{amsfonts}       % blackboard math symbols
\usepackage{nicefrac}       % compact symbols for 1/2, etc.
\usepackage{microtype}      % microtypography
\usepackage{arydshln}       % hdashline
\usepackage[margin=1in]{geometry}
\usepackage{authblk}

\usepackage{mathtools}
\usepackage{algorithm, algpseudocode}
\usepackage{graphicx}
\usepackage{wrapfig}
\usepackage{algorithm}
\usepackage{bm}
\usepackage{xcolor}
\usepackage{subcaption}
\usepackage{bbm}
\usepackage{latexsym}
\usepackage{tipa}           % for IPA notation
\DeclareMathOperator*{\argmin}{\mathrm{arg\,min}}
\DeclareMathOperator*{\argmax}{\mathrm{arg\,max}}
\DeclareMathOperator*{\softmax}{\mathrm{soft\,max}}

\newcommand{\vect}[1]{\bm{#1}}
\newcommand{\transpose}{^\top}
\newcommand{\expect}{\mathbb E}

\newcommand{\abs}[1]{\lvert #1\rvert}
\newcommand{\sabs}[1]{\left\lvert #1\right\rvert}

\DeclareMathOperator{\erf}{erf}
\DeclareMathOperator{\ncdf}{NormalCDF}

\DeclareMathOperator{\arccot}{arccot}

\newcommand{\loss}{\mathcal L}
\newcommand{\subloss}{\ell}

\begin{document}
\title{Neural circuits for dynamics-based segmentation of time series}

\date{}

\author[1]{Tiberiu Te\c sileanu}
\author[1]{Siavash Golkar}
\author[2]{Samaneh Nasiri}
\author[1,3]{Anirvan M.\ Sengupta}
\author[1,4]{Dmitri B.\ Chklovskii}
\affil[1]{Center for Computational Neuroscience, Flatiron Institute}
\affil[2]{Department of Neurology, Harvard Medical School}
\affil[3]{Department of Physics and Astronomy, Rutgers University}
\affil[4]{Neuroscience Institute, NYU Medical Center}

\maketitle

\begin{abstract}
    The brain must extract behaviorally relevant latent variables from the signals streamed by the sensory organs. Such latent variables are often encoded in the dynamics that generated the signal rather than in the specific realization of the waveform. Therefore, one problem faced by the brain is to segment time series based on underlying dynamics.
    We present two algorithms for performing this segmentation task that are biologically plausible, which we define as acting in a streaming setting and all learning rules being local. One algorithm is model-based and can be derived from an optimization problem involving a mixture of autoregressive processes. This algorithm relies on feedback in the form of a prediction error, and can also be used for forecasting future samples. In some brain regions, such as the retina, the feedback connections necessary to use the prediction error for learning are absent. For this case, we propose a second, model-free algorithm that uses a running estimate of the autocorrelation structure of the signal to perform the segmentation. We show that both algorithms do well when tasked with segmenting signals drawn from autoregressive models with piecewise-constant parameters. In particular, the segmentation accuracy is similar to that obtained from oracle-like methods in which the ground-truth parameters of the autoregressive models are known. We also test our methods on datasets generated by alternating snippets of voice recordings. We provide implementations of our algorithms at \url{https://github.com/ttesileanu/bio-time-series}.
\end{abstract}

\section{Introduction}\label{sec:intro}
Detecting changes in the environment is essential to a living organism’s survival~\citep{Koepcke2016}. To a first approximation, environmental stimuli reaching our senses are generated by switching between different dynamical systems driven by a stochastic source. This implies that the temporal dependency
structure of the stimuli, rather than their exact time evolution, contains information about the dynamics, which allows the brain to detect important changes in the environment. For example, does the AC sound different today?  Is your conversational partner’s voice sad or cheerful? Is the recent surge in electrical activity in the brain indicative of an imminent seizure? % the sudden lull in activity an indication of the presence of a predator?

%Does the AC sound different today? How does the brain detect changes in stochastic sensory data? The environmental stimuli that reach our senses are the result of a hierarchy of dynamical processes operating on multiple timescales. To a first approximation, we can view this as a stochastic source signal that drives a dynamical system, whose output is the sensory data. The sensory data is thus itself stochastic, but its temporal dependency structure contains information about the dynamics that generated it that is independent of the details of the stochastic source. Detecting changes in this temporal structure can thus shed light on changes in environmental dynamics---an essential task for living organisms~\citep{Koepcke2016}.

In this paper we develop unsupervised, biologically plausible neural architectures that segment time series data in an online fashion, clustering the underlying generating dynamics. Our focus here is different from typical applications of change-point detection in neuroscience (\emph{e.g.,}~\citep{Beck2001, Yu2007}), as we are focusing on changes in the temporal correlation structure of the data, in contrast to changes in instantaneous statistics like the mean or the variance. We are also addressing a different question compared to sequence learning or identification (\emph{e.g.,}~\citep{Brea2011,Brea2013,Memmesheimer2014}), since we aim to segment time series based on the dynamical processes that generated them, rather than the precise patterns that they contain in any given instance. Methods based on hidden Markov models (HMMs) have been used to segment spike-train data~\citep{Abeles1995,Jones2007,Mazzucato2015,Escola2011}, but these generally do not work online and are not implemented using biologically plausible circuits. Methods similar to ours, but without a neural substrate, have also been used in econometrics~\citep{ni2009self,ouyang2014neural} and for analysis of electroencephalogram (EEG) data~\citep{camilleri2015semi}.

There is also an extensive literature on probabilistic methods that perform change-point detection~\citep{Ghahramani2000, Desobry2005, Adams2007, Fox2008, Saatci2010, Roberts2013, Guo2016, Linderman2017}. Some of these algorithms work online but focus on \emph{i.i.d.}~data, ignoring temporal correlations~\citep{Desobry2005, Adams2007}. Some can handle complex autocorrelation structures but require multiple passes over the data~\citep{Ghahramani2000, Fox2008, Roberts2013}. There are also models based on Gaussian processes~\citep{Saatci2010} and recurrent neural networks~\citep{Guo2016} that handle both temporal correlations and online inference. Our work aims to provide a bridge between these methods and biologically plausible neural circuits. 

In order to build circuits that perform time-series clustering in a biologically plausible way, we require that learning occurs online and uses only local update rules. The first constraint is because biological learning and inference tend to happen in a streaming setting, with decisions taken as the sensory data is received and without the possibility of reprocessing the same data.\footnote{Hippocampal replay is a notable exception~\citep{Pavlides1989,Buhry2011}.} The second constraint reflects the fact that synaptic plasticity involves chemical processes that only have access to the local environment of the synapse. Synaptic updates thus typically depend only on the activity of pre- and post-synaptic neurons, potentially modified by a modulator, using a Hebb-like mechanism~\citep{hebb2005organization, Kusmierz2017}.

Apart from constraints like the ones above, which one would expect to hold across all brain circuits, there are also limitations specific to certain areas. For instance, the retina in mammals does not receive feedback connections from the rest of the brain~\citep{kandel2000principles}, and so the results of computations performed downstream cannot be used to inform learning in the retina. In particular, algorithms involving the calculation of prediction errors seem implausible at this level. In other parts of the brain, however, neural correlates for prediction errors have been found~\citep{Schultz1997, Cohen2007, Egner2010, Tang2018}, and thus this constraint can be lifted in those cases.

Here we show how the brain can implement time-series segmentation using two different biologically plausible architectures. If prediction error calculations are allowed, a model-based algorithm related to online $k$-means learning~\citep{sim-kmeans} can solve the task effectively. The model predicts the existence of multiple independent modules in the brain, one for each learned generating process, and one global inhibitory neuron that silences all but the module that most accurately represents the data at any given time. Because of this latter feature we will typically refer to the model-based algorithm as ``winner-take-all''. We note also that this approach is related to developments in machine learning, particularly in the field of causal learning~\citep{Bengio_representation, learning_causal,competitive_generative,unsupervised_disentangled,causality_review,competing_rnn}.

In a second, model-free approach, a running estimate of the autocorrelation structure of the signal is clustered using a non-negative similarity-matching algorithm~\citep{Hu2014, Minden2018}. By employing a metric that focuses on the similarity structure instead of encoding error, the model-free approach provides an architecture that does not require any feedback connections. This approach can therefore model brain areas in which such feedback is not available. The distinction between our model-based and model-free algorithms is similar to the difference between parametric and non-parametric models~\citep{Deng97}.

In the following sections, we will formally define the segmentation task; we will then introduce the winner-take-all algorithm and the autocorrelation-based algorithms together with their biological implementations; next we will compare their segmentation performance with each other and with an oracle-like method with roots in control theory; and we will end with a summary and discussion of future work.

\section{Piecewise stationary autoregressive dynamics}\label{sec:problem}

\begin{figure}[htb]
    \centering
    \includegraphics[width=5.76in]{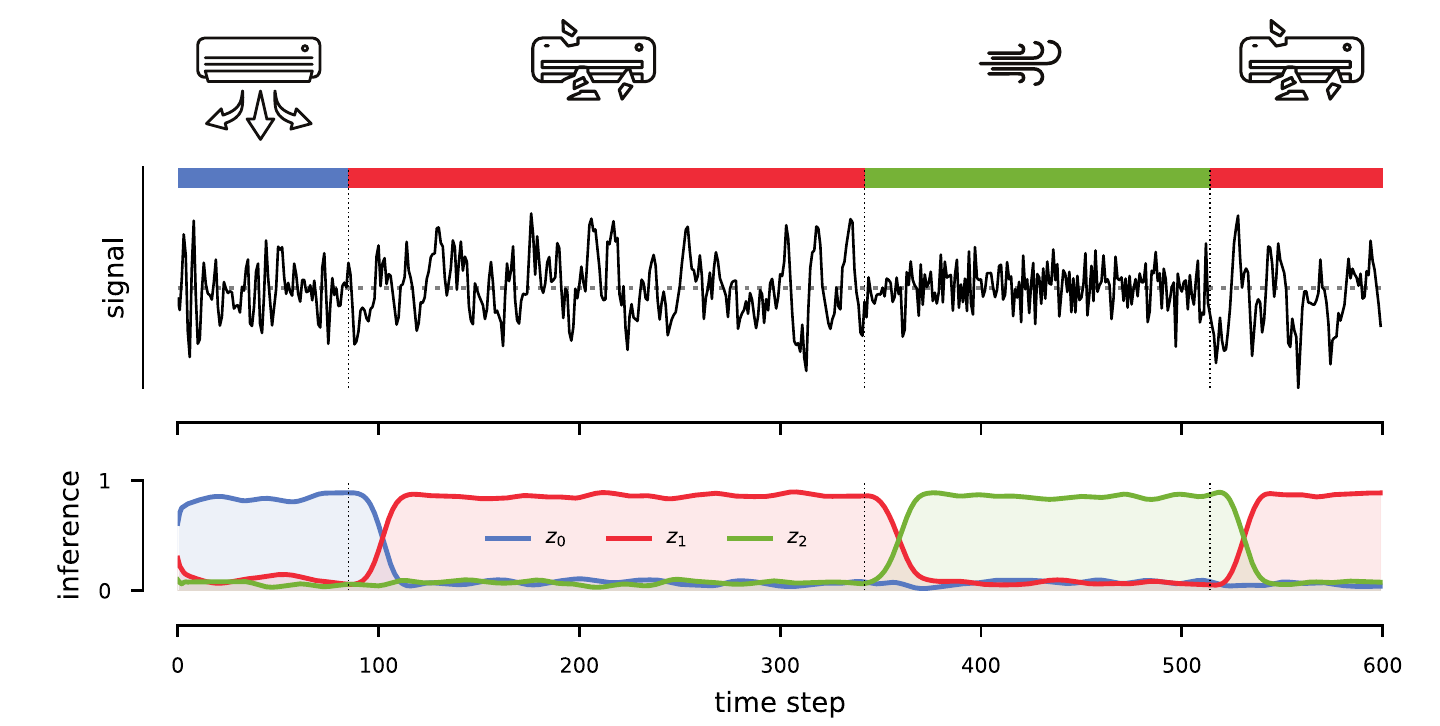}
    \caption{Sketch of the inference task. A signal (black line in top panel) is generated by alternating processes---\emph{e.g.,} an intact AC, a broken AC, a gust of wind, as shown in the icons on the top, and also indicated by the colored ribbon between the icons and the signal line. The segmentation task amounts to identifying the transitions and clustering the sources (bottom panel). The $z_k$ curves sketched in the bottom panel perform a soft clustering. % Note that learning is unsupervised, so labels can be mismatched between inference and ground truth (\emph{e.g.,} broken AC is clustered as red in the top panel, green in the bottom one).
    We also sketched a delay in recognizing each transition, which is inherent in an online setting.}
    \label{fig:example_signal}
    % https://thenounproject.com/term/ac/3674454/
    % https://thenounproject.com/term/broken-ac/2268823/
    % https://thenounproject.com/search/?q=wind&i=1244736
\end{figure}
More formally, we consider time series data $y(t)$ whose structure at any given time $t$ is induced by one of a number of different stationary autoregressive (AR) processes (see Figure~\ref{fig:example_signal}). Mathematically,
\begin{equation}
    y(t) = \begin{cases}
        w_{11} y(t-1) + \dotsb + w_{1p} y(t-p) + \epsilon(t) \,, & \text{if $\hat z(t) = 1$,}\\
        \hfill \vdots \hfill & \\
        w_{M1} y(t-1) + \dotsb + w_{Mp} y(t-p) + \epsilon(t) \,, & \text{if $\hat z(t) = M$,}
    \end{cases}
\end{equation}
where $\hat z(t)$ indicates the process that generated the sample at time $t$, and $\epsilon(t) \sim \mathcal N(0, \sigma^2)$ is white Gaussian noise. We can write this more compactly as
\begin{equation}
    \label{eq:regression}
    y(t) = \sum_{k=1}^M z_k(t) \Bigl[\vect w_k\transpose \vect x(t) + \epsilon(t)\Bigr]\,,
\end{equation}
where we introduced the notation $\vect x(t)$ for the \emph{lag vector} with components
\begin{equation}
    x_i(t) = y(t-i)\,, \qquad i\in\{1, \dotsc, p\}\,,
\end{equation}
and used a one-hot encoding for the latent-state variable $\hat z_k(t)$,
\begin{equation}
    z_k(t) = \begin{cases}
        1 & \text{if $\hat z(t) = k$,}\\
        0 & \text{else.}
    \end{cases}
\end{equation}

Our aim is thus to develop a biologically plausible mechanism that assigns each sample to a particular generative process (segmentation) and, in the model-based case, infer the process parameters (system identification). More specifically, this amounts to inferring the latent states $z_k(t)$ for the segmentation task and estimating the coefficients $\vect w_k$ for the system identification task.

There are several generalizations that we will not address in detail, but are straightforward to implement: handling processes with non-zero mean; allowing for more complex dependencies on past samples; and working with multidimensional time series. See Appendix~\ref{sec:app:generalizations}. In this work, we focus on the special case described in eq.~\eqref{eq:regression}.

\section{Model-based, winner-take-all algorithm}\label{sec:biowta}

\subsection{Basic framework}
A natural approach for solving both the segmentation and system-identification tasks outlined above is to find the latent states $z_k(t)$ and AR coefficients $\vect w_k$ that minimize the discrepancy between the values of the signal predicted from eq.~\eqref{eq:regression} and the actual observed values $y(t)$,
\begin{equation}
    \min_{\vect w_k} \min_{\vect z} \frac 1 {2\sigma^2} \sum_t \sum_{k=1}^M z_k(t) \bigl\lvert y(t) - \vect w_k\transpose \vect x(t)\bigr\rvert^2 \quad \text{such that $z_k(t)$ is one-hot,}
    \label{eq:bio_objective_original}
\end{equation}
where $\sigma$ is the standard deviation of the noise $\epsilon(t)$ from eq.~\eqref{eq:regression}. Note that this has a probabilistic interpretation: it is a maximum likelihood method based on the assumption that the noise $\epsilon(t)$ is normally distributed.

The optimal $z_k(t)$ values at fixed $\vect w_k$ are given by the following expression (see Appendix~\ref{sec:app:biowta_derivation}):
\begin{equation}
    \label{eq:bio_z_update}
    z_k(t) = \delta_{kk^*(t)}\,,\qquad k^*(t) = \argmax_k \, -\frac 1 {2\sigma^2} \bigl\lvert y(t) - \vect w_k\transpose \vect x(t)\bigr\rvert^2\,,
\end{equation}
where $\delta_{kk^*} = 1$ if $k=k^*$ and $0$ otherwise is the Kronecker delta. Intuitively, the best estimate for the latent state at time $t$ is the one that produces the lowest prediction error. This depends both on the current estimate for the model coefficients $\vect w_k$ and on the recent history of the signal, represented by the lag vector $\vect x(t)$.

The optimal latent state assignments $z_k(t)$ and the optimal process parameters $\vect w_k$ depend on each other, which means that a full solution to the optimization problem~\eqref{eq:bio_objective_original} requires iteratively re-evaluating all the latent state variables $z_k(1), \dotsc, z_k(t)$ and the coefficients $\vect w_k$. This is analogous to Lloyd's iterative solution for $k$-means clustering~\citep{Lloyd1982}, but is unsuitable for an online algorithm where samples are presented one at a time and are generally not stored in memory.

To obtain an online approximation, we assume that the latent-state estimates $z_k(t)$ do not change once they are made. We thus apply eq.~\eqref{eq:bio_z_update} only once for each time step $t$, and we then use stochastic gradient descent to update the coefficients $\vect w_k$ given a new sample. This yields:
\begin{equation}
    \label{eq:weight_updates}
    \vect w_k(t+1) = \vect w_k(t) + \eta_w z_k(t) \vect x(t) \bigl[ y(t)  - \vect x(t)\transpose \vect w_k(t)\bigr]\,,
\end{equation}
where $\eta_w$ is a learning rate.

We call the algorithm based on eqns.~\eqref{eq:bio_z_update} and~\eqref{eq:weight_updates} ``winner-take-all'' because only one $z_k^*(t)$ is non-zero and only the weights associated with the inferred latent state, $\vect w_{k^*(t)}$, are updated at each step. %This is because $z_k(t)$ is zero for every other state, $z_{k\ne k^*(t)}(t) = 0$. 
Below we will relax this condition by softening the clustering and allowing several $z_k(t)$ to be non-zero, but it will generally hold true that the weight updates are strongest for the process that yields the best prediction for the sample. We will therefore sometimes refer to the algorithm as ``soft'' or ``enhanced'' winner-take-all.

\subsection{Enhancements to the basic method}
\label{sec:biowta_enhancements}
\paragraph{Soft clustering.}~Instead of forcing the latent-state vector $\vect z$ to be one-hot, as in eq.~\eqref{eq:bio_z_update}, we can soften the clustering by employing a $\softmax$ nonlinearity instead,
\begin{equation}
    \label{eq:bio_z_update_soft}
    z_k(t) = \softmax{}_T \left\{-\frac 1 {2\sigma^2} \bigl\lvert y(t) - \vect w_k(t)\transpose \vect x(t)\bigr\rvert^2\right\}\,,
\end{equation}
where
\begin{equation}
    \softmax{}_T \Delta_k = \frac {e^{\Delta_k/T}} {\sum_{k'} e^{\Delta_{k'}/T}}\,.
\end{equation}
Here $T$ is a ``temperature'' parameter controlling the softness of the clustering. In the $T\to 0$ limit, this reduces to the $\argmax$ solution.

This is equivalent to the following change in the objective function (\emph{cf.}~eq.~\eqref{eq:bio_objective_original}; see derivation in Appendix~\ref{sec:app:biowta_derivation}):
\begin{equation}
    \min_{\vect w_k} \min_{\substack{\vect z\geq0,\\\sum z_k = 1}} \frac 1 {2\sigma^2} \sum_t \sum_{k=1}^M z_k(t) \bigl\lvert y(t) - \vect w_k\transpose \vect x(t)\bigr\rvert^2 \textcolor{blue}{\,+\, T \left[\sum_{k=1}^M z_k \log z_k-1\right]}\,.
    \label{eq:bio_objective_soft}
\end{equation}

\paragraph{Persistence of latent states: penalizing transitions.}~In many realistic situations, the latent states exhibit some level of persistence: if the signal was generated by model $k^*(t)$ at time $t$, we can assume that the signal at time $t+1$ will likely be generated by the same model. Assuming persistence helps to avoid spurious switches in the inferred latent states that are due to noise. The downside is that actual switches in the state are more likely to be dismissed as noise.

One way to encourage persistence of the inferred latent states is to add a pairwise interaction term to the loss function:
\begin{equation}
    \min_{\vect w_k} \min_{\vect z\geq0} \frac 1 {2\sigma^2} \sum_t \sum_{k=1}^M z_k(t) \bigl\lvert y(t) - \vect w_k\transpose \vect x(t)\bigr\rvert^2 \textcolor{blue}{\,-\, J \sum_k \sum_t z_k(t-1) z_k(t)}\,,
    \label{eq:wta_objective_with_continuity}
\end{equation}
where $J$ controls the strength of the persistence correction. The extra term can be seen as a regularizer, or equivalently, as imposing a prior on the structure of the latent states. This latter interpretation is related to the fact mentioned above that the optimization from eq.~\eqref{eq:bio_objective_original} is the maximum-likelihood solution for the generative model from eq.~\eqref{eq:regression}. In this language, adding the persistence correction turns a maximum-likelihood technique into a maximum \emph{a posteriori} (MAP) approach.

In the online algorithm, this regularization has the effect of penalizing states that are different at time $t$ from the state at time $t-1$ by adding a term proportional to $J$ to their squared prediction errors. Mathematically, eq.~\eqref{eq:bio_z_update} is replaced by
\begin{equation}
    \label{eq:bio_z_update_with_continuity}
    z_k(t) = \delta_{kk^*(t)}\,,\qquad k^*(t) = \argmax_k \, \left\{-\frac 1 {2\sigma^2} \bigl\lvert y(t) - \vect w_k(t)\transpose \vect x(t)\bigr\rvert^2 \textcolor{blue}{\,+\, J z_k(t-1)}\right\}\,.
\end{equation}

\paragraph{Averaging the squared error.}~A different approach that combines the signal across several consecutive samples is to replace the instantaneous squared prediction error $\abs{y(t) - \vect w_k(t)\transpose \vect x(t)}^2$ in eq.~\eqref{eq:bio_z_update_with_continuity} with a time-average,
\begin{equation}
    \label{eq:bio_z_update_with_averaging}
    \begin{split}
        z_k(t) &= \delta_{kk^*(t)}\,,\\
        k^*(t) &= \argmax_k \, -\frac {\textcolor{blue}{\eta_\Delta}} {2\sigma^2} \Bigl\{\abs{y(t) - \vect w_k(t)\transpose \vect x(t)}^2\\
        &\qquad +\textcolor{blue}{(1 - \eta_\Delta) \abs{y(t-1) - \vect w_k(t-1)\transpose \vect x(t-1)}^2 }\\
        &\qquad +\textcolor{blue}{(1 - \eta_\Delta)^2 \abs{y(t-2) - \vect w_k(t-2)\transpose \vect x(t-2)}^2}\\
        &\qquad \textcolor{blue}{{}+ \dotsb}\Bigr\}\\
        &\equiv \argmax_k \, -\frac 1 {2\sigma^2} \Delta_k(t)\,,
    \end{split}
\end{equation}
where we used the notation $\Delta_k(t)$ for the exponential moving average (EMA) of the squared prediction error with smoothing factor $\eta_\Delta$. This can be calculated online using
\begin{equation}
    \Delta_k(t+1) = \eta_\Delta \abs{y(t) - \vect w_k(t)\transpose \vect x(t)}^2 + (1 - \eta_\Delta)\Delta_k(t)\,.
    \label{eq:low_pass_filtered_error}
\end{equation}
Averaging the squared error mitigates the effect of noise on latent-state inference much like the penalty on state transitions described above does.

\paragraph{Final expression for latent-state estimates.}~Combining all the techniques in this section, we obtain the following expression for inferring the identity of the latent states:
\begin{equation}
    \label{eq:bio_z_softmax}
    z_k(t) = \softmax{}_T \, \left\{-\frac 1 {2\sigma^2} \Delta_k(t) + J z_k(t-1)\right\}\,.
\end{equation}
% This latter form is the one we will use throughout the rest of the paper.

\subsection{Biologically plausible implementation}
% text width in figure: 5.76884in
\begin{figure}[!tbh]
    \centering
    \includegraphics{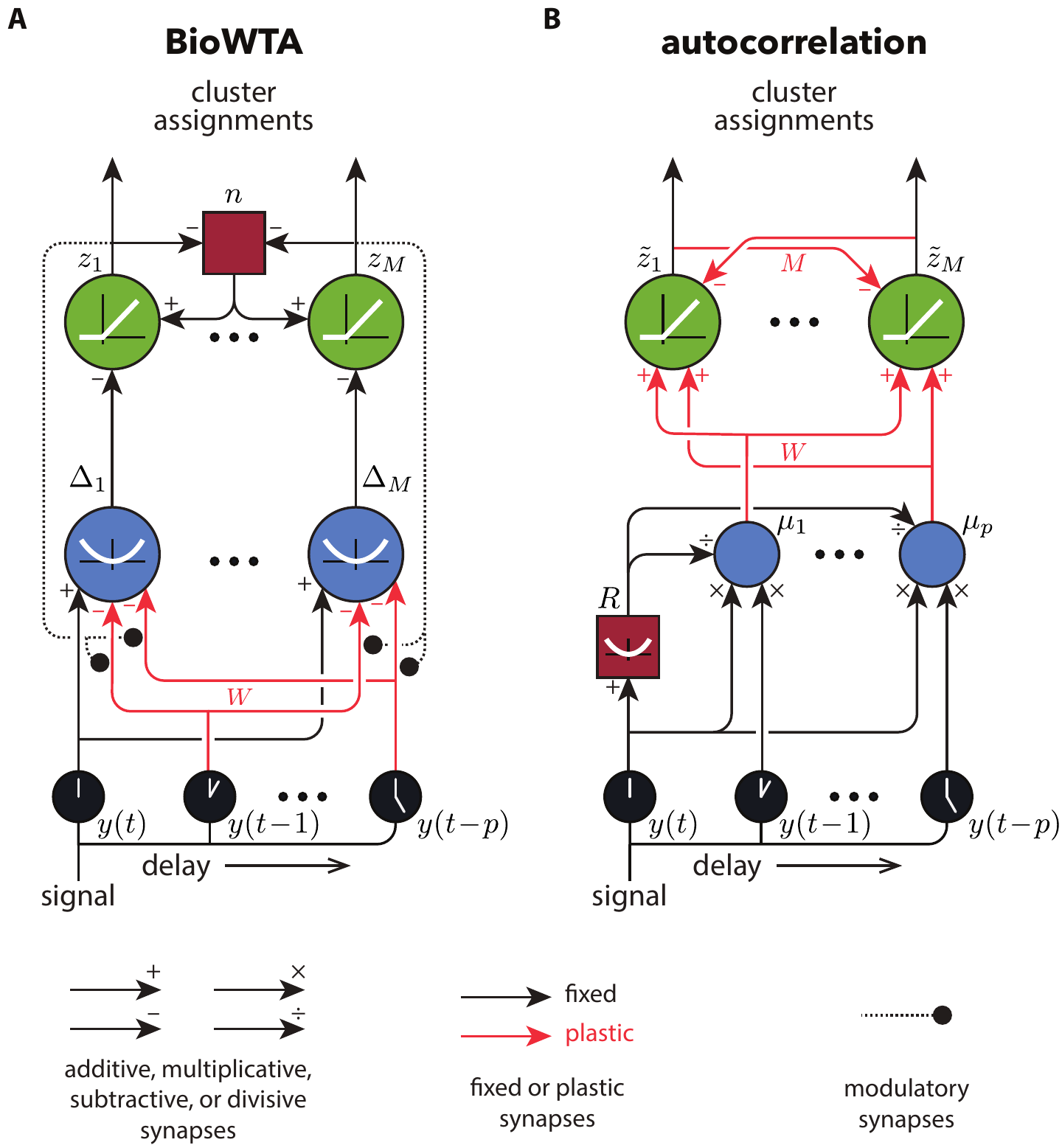}
    \caption{Biological implementations of our algorithms. (A) The model-based winner-take-all algorithm. (B) The model-free autocorrelation-based algorithm. The neurons are linear unless an activation function is shown. The neurons in blue are leaky integrators with timescales related to the appropriate learning rates in the models (see text); the other neurons are assumed to respond instantaneously.}
    \label{fig:bio_circuits}
\end{figure}

\begin{algorithm}[!htb]
    \caption{Biological winner-take-all method}
    \label{alg:bio-WTA}
    \begin{algorithmic}
        \Function{ProcessSample}{$\vect x$, $y$, $z_k^\text{prev}$}
        \State $\Delta_k \gets (1-\eta_\Delta) \Delta_k + \eta_\Delta \sabs{y(t) - \vect w_k\transpose \vect x}^2\,.$\Comment{averaged reconstruction error}
        \Repeat\Comment{output-neuron dynamics}
            \State $z_k \gets z_k - \eta_z \Bigl[\frac 1 {2\sigma^2} \Delta_k - J z_k^\text{prev} + n + T \log z_k\Bigr]\,,$
            \State $n \gets n + \eta_n \Bigl( \sum_k z_k -1\Bigr)\,.$
        \Until{convergence}
        \State $\vect w_k \gets \vect w_k +  \eta_w z_k \vect x \bigl[y  - \vect w_k\transpose\vect x\bigr]\,.$\Comment{synaptic updates}
        \State \Return $z_k\,.$
        \EndFunction
    \end{algorithmic}
\end{algorithm}

We now construct a circuit that can implement the winner-take-all algorithm defined in eqns.~\eqref{eq:bio_z_softmax} and~\eqref{eq:weight_updates} in a biologically plausible way. The key observation is that the latent state $z_k(t)$ can be obtained from the following optimization problem (see Appendix~\ref{sec:app:biowta_derivation}):
\begin{equation}
    \min_{\vect z\geq0} \max_n \sum_{k=1}^M z_k(t) \left[\frac 1 {2\sigma^2} \Delta_k(t) - J z_k(t-1) + T \bigl(\log z_k(t) - 1\bigr) - n(t)\right] + n(t)\,,
    \label{eq:wta_objective_with_continuity_and_averaging}
\end{equation}
where $\Delta_k(t)$ is the EMA of the squared prediction error (eq.~\eqref{eq:low_pass_filtered_error}) and $n(t)$ is a Lagrange multiplier enforcing the constraint $\sum_k z_k(t) = 1$. Similar to~\citep{sim-kmeans}, we can solve this min-max optimization objective with gradient descent-ascent dynamics:
\begin{equation}
    \label{eq:neural_dynamics}
    \begin{split}
        \dot z_k(t) &= \eta_z \left(-\frac 1 {2\sigma^2}\Delta_k(t) + J z_k(t-1) + n(t) - T \log z_k(t)\right)\,,\\
        \dot n(t) &= \eta_n \Bigl(1 - \sum_k z_k(t)\Bigr)\,.
    \end{split}
\end{equation}
Note that here $t$ refers to the sample index, while the dynamics happens on a fast timescale that must achieve convergence before the next sample can be processed.
% where $\mu$ is an iteration index and $[a]_+\equiv \max(a,0)$ is the linear rectification function. The rectification appears here because of the non-negativity constraint on the latent-state indicator variables $z_k(t)$. The final value of the $z$ variables is set at convergence, $z_k(t) \equiv z^\infty_k(t)$. A good way to initialize the iteration is using the values from the previous time step, $z^0_k(t) = z_k(t-1)$, $n^0_k(t) = n_k(t-1)$.

Running the neural program above until convergence recovers the $\softmax$ solution of eq.~\eqref{eq:bio_z_update_soft}. The $T\log z_k(t)$ term enforces the non-negativity constraint on $z_k(t)$ by going to negative infinity as $z_k(t) \to 0$.

Combining eq.~\eqref{eq:neural_dynamics} with the synaptic plasticity rule from eq.~\eqref{eq:weight_updates},
\begin{equation}
    \label{eq:weight_updates_repeated}
    \vect w_k(t+1) = \vect w_k(t) + \eta_w z_k(t) \vect x(t) \bigl[ y(t)  - \vect x(t)\transpose \vect w_k(t)\bigr]\,,
\end{equation}
yields the basis of our biological winner-take-all neural circuit. The method is summarized in Algorithm~\ref{alg:bio-WTA}, and the resulting circuit, providing a biological implementation of a neural attention mechanism modulated via competition, is sketched in Figure~\ref{fig:bio_circuits}A .

The neurons labeled $z_k$ in Figure~\ref{fig:bio_circuits}A represent the cluster assignments and compete with each other via the interaction with the interneuron $n$. The ``winning'' clusters get their parameters updated following a three-factor Hebbian learning rule~\citep{Kusmierz2017}, eq.~\eqref{eq:weight_updates_repeated}, at the $x\to \Delta$ synapses, where the outputs from the $z_k$ neurons are used as modulators. The circuit uses leaky integrator neurons with a quadratic nonlinearity ($\Delta_k$) to estimate the average squared error from eq.~\eqref{eq:low_pass_filtered_error}. These neurons project to the output neurons ($z_k$) that implement the $\softmax$ function in conjunction with the normalizing ($n$) neuron, as in eq.~\eqref{eq:neural_dynamics}. % Note that the prediction error $\Delta_k(t) \sim y(t) - \vect w_k\transpose \vect x(t)$ can be either positive or negative while neuronal activations are positive. We assume that a biological circuit would solve this by, for instance, using two rectifying neurons or two dendritic branches within the same neuron to represent the positive and negative activations. % Mathematically, this amounts to taking advantage of the identity $\bigl\lvert y(t) - \vect w_k\transpose \vect x(t)\bigr\rvert^2 = \bigl(y(t)-\vect w_k\transpose \vect x(t)\bigr)_+^2 + (-y(t) + \vect w_k\transpose \vect x(t)\bigr)_+^2$, where $(x)_+ = x$ if $x > 0$ and 0 otherwise.

The BioWTA algorithm thus relies on the selection of one (or a few) clusters, and updating the corresponding weights using a Hebb-like rule. This is achieved in a biologically plausible way by using recurrent connectivity between the excitatory $z$ neurons and the inhibitory neuron $n$ (see Figure~\ref{fig:bio_circuits}A) to silence all but the most strongly active $z$ neurons. This information can be fed back to the $y\to \Delta$ synapses using a neuromodulator like dopamine, which has been shown to modulate synaptic plasticity in certain contexts~\citep{Gurden1999,Gurden2000,Navakkode2017}. The slow dynamics expected from such a mechanism might be an asset for our model, since it would contribute to the smoothing of the error signal which ensures continuity of the latent state estimates (similar to eq.~\eqref{eq:bio_z_update_with_averaging}). Our model of course includes only a very simple interaction between synaptic plasticity and the modulator, and it would be interesting to see what happens when more realistic details are added to the circuit. This, however, is beyond the scope of this paper.

\section{Model-free, autocorrelation-based algorithm}

\begin{algorithm}[tbh]
    \caption{Biological autocorrelation method}
    \label{alg:bio-xcorr}
    \begin{algorithmic}
        \Function{ProcessSample}{$\vect x$, $y$}
        \State $R \gets R + \eta_R (y^2 - R)\,.$ \Comment{autocorrelation update}
        \State $\vect \mu \gets \vect \mu + \eta_\mu \bigl(y\vect x / R - \vect \mu\bigr)\,.$\bigskip
        \State $M_d \gets \text{diagonal part of $M$}\,.$\Comment{output-neuron dynamics}
        \State $M_o \gets \text{off-diagonal part of $M$}\,.$
        \State $\tilde {\vect z}_d \gets M_d^{-1} (W \vect \mu)\,.$
        \State $\tilde {\vect z} \gets \tilde {\vect z}_d - M_d^{-1} (M_o \tilde {\vect z}_d)\,.$\bigskip
        \State $W \gets W + \alpha \, [\tilde {\vect z} \vect \mu\transpose - W]\,.$\Comment{synaptic updates}
        \State $M \gets M + \tau^{-1} \alpha \, [\tilde {\vect z} \tilde {\vect z}\transpose - M]\,.$
        \State \Return $\tilde z_k\,.$
        \EndFunction
    \end{algorithmic}
\end{algorithm}

In this section, we propose a network that operates without an explicit error calculation, in contrast to the winner-take-all circuit which relies on an estimate of the prediction error for both inference and learning. To do so, we combine a running estimate of the autocorrelation structure of the signal with a biologically plausible clustering algorithm.

The key observation is that the dynamical characteristics of a signal can be summarized through its autocorrelation structure. Indeed, the coefficients defining an autoregressive process are related to the autocorrelation function of the signal it generates through the Yule-Walker equations~\citep{shumway2000time}, although the precise relationship is not important here. We summarize the autocorrelation structure using a $p$-dimensional vector $\vect \mu$ with components
\begin{equation}
    \label{eq:acorr_vec_meaning}
    \begin{split}
        \mu_k &= \frac 1R \expect[y(t) y(t + k)]\,, \qquad \text{where $R$ is the variance,}\\
        R &= \expect[y(t)^2]\,.
    \end{split}
\end{equation}
Time-series segmentation then reduces to calculating short-time estimates of the autocorrelation vectors $\vect \mu(t)$, and clustering the vectors obtained at different moments in time.

The following set of update rules can be used to estimate the autocorrelation structure in a streaming setting:
\begin{equation}
    \label{eq:acorrOnline}
    \begin{split}
        \Delta R(t) &= \eta_R \bigl(y(t)^2 -R(t)\bigr)\,, \\
        \Delta \vect \mu(t) &= \eta_\mu \left[\frac 1 {R(t)} y(t) \vect x(t) - \vect \mu(t)\right]\,,
    \end{split}
\end{equation}
where $\eta_R$ and $\eta_\mu$ are learning rates. Note that, as in the winner-take-all algorithm, we are assuming that the input signal has mean zero. If it does not, a simple adaptation mechanism could be used to subtract the mean.

To cluster the vectors $\vect \mu(t)$ that summarize the dynamics at time $t$, we use non-negative similarity matching (NSM), an online algorithm that admits a simple neural interpretation~\citep{Hu2014, Minden2018}. The NSM algorithm is based on the idea that signals with similar autocorrelation structure should be mapped to similar outputs,
\begin{equation}
    \label{eq:nsm_loss}
    \argmin_{\tilde {\vect z}(t) \ge 0} \sum_{t, t'} \sabs{\vect \mu(t)\transpose \vect \mu(t') - \tilde{\vect z}(t)\transpose \tilde{\vect z}(t')}^2\,,
\end{equation}
where we force the outputs to be non-negative, $\tilde z_k(t) \ge 0$. With this constraint, the outputs of the network perform soft clustering~\citep{pehlevan2014hebbian}, such that $\tilde z_k(t)$ can act as an indicator function for whether the $k^\text{th}$ generating process is responsible for the output at time $t$.

Note that unlike in the case of BioWTA, the outputs $\tilde z_k(t)$ from the autocorrelation-based algorithm do not in general sum to 1. This is why we use a slightly different notation here, $\tilde z_k(t)$ instead of $z_k(t)$, for the outputs. We can still recover the best guess for the latent state at time $t$, $z_k(t)$, by finding the largest $\tilde z_k(t)$:
\begin{equation}
    z_k(t) = \delta_{kk^*(t)}\,, \text{with }k^*(t) = \argmax_k \tilde z_k(t)\,.
\end{equation}
The optimization from eq.~\eqref{eq:nsm_loss} can be implemented using the following equations~\citep{Minden2018}:
\begin{equation}
    \label{eq:nsm_dynamics}
    \begin{split}
        \tilde {\vect z}(t) &= \bigl[M(t)^{-1} W(t) \vect \mu(t)\bigr]_+\,,\\
        W(t + 1) &= W(t) + \alpha \bigl[\tilde {\vect z}(t) \vect \mu(t)\transpose - W\bigr]\,,\\
        M(t + 1) &= M(t) + \tau^{-1} \alpha \bigl[\tilde {\vect z} (t) \tilde {\vect z}(t)\transpose - M\bigr]\,,
    \end{split}
\end{equation}
where $\alpha$ and $\tau^{-1} \alpha$ are learning rates and $[u]_+$ denotes a rectifying nonlinearity. Note how the synaptic weights $W$ and $M$ undergo Hebb-like dynamics. As in~\citep{Minden2018}, in practice we use a more biologically plausible two-step approximation instead of calculating the inverse $M(t)^{-1}$ in the equation for $\tilde {\vect z}(t)$; see Algorithm~\ref{alg:bio-xcorr}.

A downside of the autocorrelation approach is that the coefficients $\vect w_k$ describing the generating processes are difficult to recover. Information about them is in principle contained in the weight matrices $M$ and $W$, which encode information about the autocorrelation structure characteristic of each process. However, obtaining the AR coefficients from the weight matrices is a non-trivial task that our circuit does not perform.

We note also that the model-free algorithm implicitly assumes a level of persistence of the latent states because of the updating rules from eq.~\eqref{eq:acorrOnline}: the system needs a number of samples of order $1 / \eta_\mu$ before it can detect a change in the autocorrelation structure, and so transitions happening on timescales faster than this will typically go undetected. The learning rate $\eta_\mu$ in the autocorrelation model thus plays a similar role as the $\eta_\Delta$ parameter used in the averaging step of the BioWTA algorithm, eq.~\eqref{eq:low_pass_filtered_error}.

The overall dynamics of the autocorrelation model combined with the clustering model can be represented using the neural architecture shown in Figure~\ref{fig:bio_circuits}B. It is assumed that a quadratic nonlinearity from the signal neurons $y$ to the interneuron $R$ implements the necessary squaring operation, with leaky integration responsible for averaging over recent samples. Inhibitory connections from the interneuron to the autocorrelation neurons $\mu_k$ perform the divisive normalization from eq.~\eqref{eq:acorrOnline}. The multiplications needed for updating the covariances can be the result of the synergistic effects of simultaneous spikes reaching the same neuron~\citep{bugmann1991summation}. A rectifying nonlinearity ensures the non-negativity of the outputs, and the synaptic updates from eq.~\eqref{eq:nsm_dynamics} follow Hebbian and anti-Hebbian rules for the feedforward and lateral connections, respectively.

We note that the autocorrelation-based algorithm does not depend on the modulation of synaptic plasticity required by the BioWTA circuit, and instead uses classic Hebbian learning. The all-to-all lateral inhibition in the autocorrelation circuit ensures that only one (or a few) of the output neurons dominate(s) at any given time (see Figure~\ref{fig:bio_circuits}B).

\section{Numerical results}
\label{sec:experiments}

\begin{table}[htb]
    \centering
    \caption{Performance measures for our algorithms. Results are summarized over 100 runs, each run using a different 200,000-sample long signal generated using alternating AR(3) processes. The same 100 signals were used across the different algorithms. The plain BioWTA algorithm assumes hard clustering and no relation between latent states $z(t)$ at different times. The enhanced BioWTA algorithm uses soft clustering (eq.~\eqref{eq:bio_z_update_soft}) and the persistence correction (eq.~\eqref{eq:bio_z_update_with_continuity}) described in the text. The cepstral algorithm assumes that the ground-truth AR coefficients are known and uses a running estimate of a cepstral norm to identify the generating process at each time (see text and Appendix~\ref{sec:app:cepstral_oracle}).}\bigskip
    \label{tab:accuracy_summary}
    \begin{tabular}{lrrrr}
\toprule
{} &  autocorr. &  plain BioWTA &  enh. BioWTA &  cepstral \\
\midrule
Mean seg. score         &       0.75 &          0.72 &         0.88 &      0.89 \\
Fraction well-segmented         &       0.40 &          0.17 &         0.73 &      0.79 \\
Seg. score of bottom 5\% &       0.52 &          0.54 &         0.62 &      0.72 \\
Mean weight error               &        --- &          1.04 &         0.83 &       --- \\
Mean convergence time           &     620 &       16320 &      12700 &      --- \\
\bottomrule
\end{tabular}
\end{table}

We consider several ways to assess the effectiveness of our algorithms: segmentation accuracy; speed of convergence; and accuracy of system identification. We measure segmentation accuracy using a score equal to the fraction of time steps for which the inferred latent state is equal to the ground truth, up to a permutation.\footnote{Since ours is an unsupervised learning task, the ordering of models in the simulations can be different from the ground-truth ordering. We choose the mapping between simulation labels and ground-truth labels that maximizes the segmentation score. This implies that the segmentation score is always $\ge 1/M$, where $M$ is the number of clusters.} We measure the speed of convergence by the number of steps needed to reach 90\% of the final segmentation score. And we measure the accuracy of system identification by the root-mean-squared difference between the learned AR coefficients and the ground-truth coefficients, normalized by the size of the difference between the ground-truth coefficients. See Appendix~\ref{sec:app:accuracy_measures} for details.

We use artificially generated time series to test our algorithms, as this gives us access to unambiguous ground-truth data (but see also section~\ref{sec:vowels}). More specifically, we generate time series data by stochastically alternating several AR generative processes, themselves having coefficients that are chosen randomly for each signal. The switch between processes is governed by a semi-Markov model, ensuring a given minimum dwell time in every state. Beyond that minimum time, we use a constant probability of switching at each step, which is chosen to achieve a given average dwell time. To source the signals, we use a constant noise scale ($\epsilon(t)$ in eq.~\eqref{eq:regression}), and we normalize the entire signal's variance to $1$ before feeding it into the segmentation algorithms. See Appendix~\ref{sec:app:signal_generation} for details.

In the simulations below, unless otherwise indicated, we use signals 200,000-samples long generated from two alternating AR(3) processes, each with a minimum dwell time of 50 steps and an average of 100. Typically only a fraction of the samples are needed for learning.

Table~\ref{tab:accuracy_summary} summarizes the performance of our algorithms on a few different metrics, and compares it to an oracle-like cepstral method rooted in control theory. The sections below provide some detail and context for these results.

\subsection{Winner-take-all algorithm is highly accurate}
\label{sec:biowta_is_accurate}
\begin{figure}[tbh]
    \centering
    \includegraphics{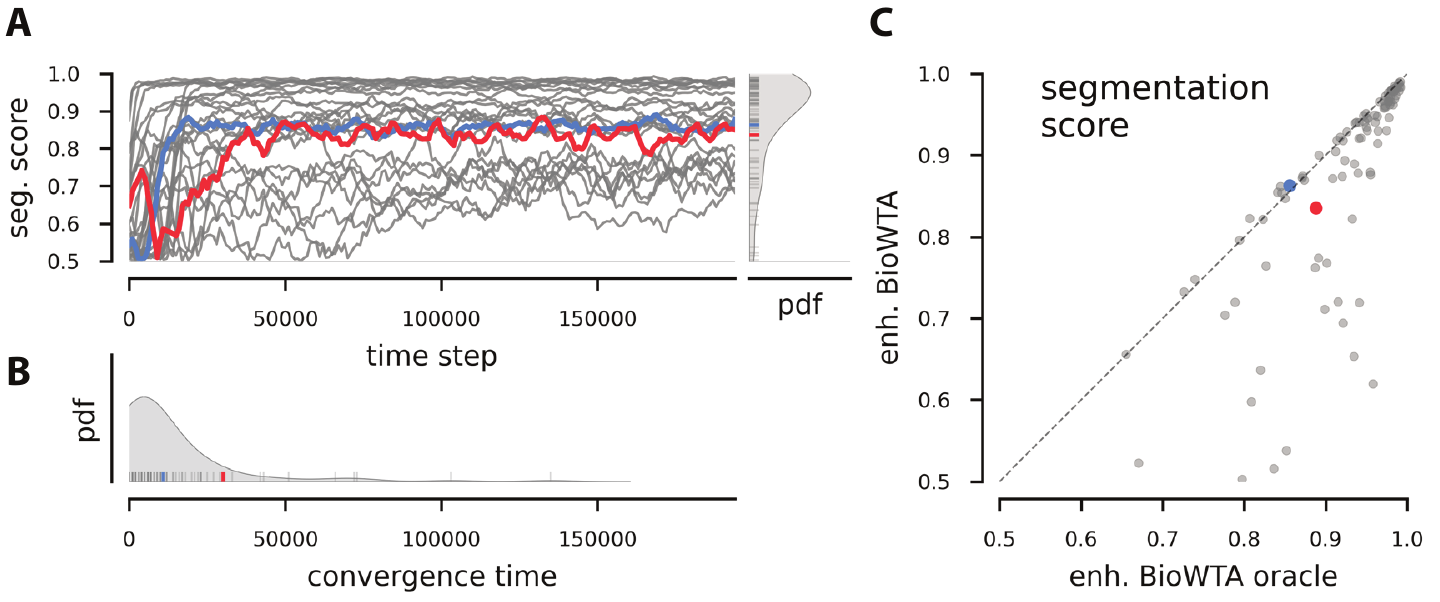}
    \caption{Segmentation accuracy for our BioWTA algorithm. (A) Rolling segmentation score for a subset of 10 of the 100 simulated runs. The blue and red traces single out runs that are used in subsequent figures. A kernel-density estimate of the distribution of segmentation scores for all 100 runs is shown on the right. (B) Kernel-density estimate of the convergence times for all 100 runs. Convergence is defined as reaching a segmentation score that is at least 90\% of the final score. (C) Comparison of segmentation scores from our BioWTA learning algorithm (on the $y$-axis) with the scores obtained from a BioWTA ``oracle''---an otherwise identical inference procedure in which the weights are kept fixed and equal to the ground-truth values.}
    \label{fig:rolling_accuracy_biowta}
\end{figure}
To test the results of the winner-take-all algorithm, we first had to choose values for the learning rates $\eta_w$, $\eta_\Delta$, the persistence parameter $J$, and the ``temperature'' from the $\softmax$ function, $T$. We did this by generating 200 random signals and running the simulation on these signals for 2000 randomly generated hyperparameters choices. We then selected parameter values that maximized the fraction of successful runs (defined as runs reaching a segmentation score of at least 85\%). We obtained the best performance by using soft clustering $T>0$ and a non-zero persistence parameter $J$, but no averaging, $\eta_\Delta = 1$. We call this the ``enhanced'' BioWTA algorithm. See Appendix~\ref{sec:app:hyperopt} for details.

Figure~\ref{fig:rolling_accuracy_biowta} shows the performance of this enhanced winner-take-all algorithm on a new batch of 100 simulated time series. We find that segmentation is typically very accurate, reaching a median score after learning of 93\%, with almost three quarters of runs exhibiting scores over 85\%. Learning is relatively fast, too: two thirds of runs converge to 90\% of their final segmentation scores in less than 10,000 time steps.

An interesting aspect of the segmentation performance is that in some cases it is not symmetric: one of the processes is misidentified more often than the other one. This happens partly due to stochastic effects such as differences in initial conditions or differences in the sequence of change points. Interestingly, though, there is also a systematic component: for certain pairs of AR(3) processes, one of them is always harder to identify, for reasons that are not immediately clear (see Appendix~\ref{sec:app:asymmetry} and Figure~\ref{fig:asymmetry_results}).

\subsection{Model coefficients are also learned by winner-take-all algorithm}
\begin{figure}[!tbh]
    \centering
    \includegraphics[width=5.76in]{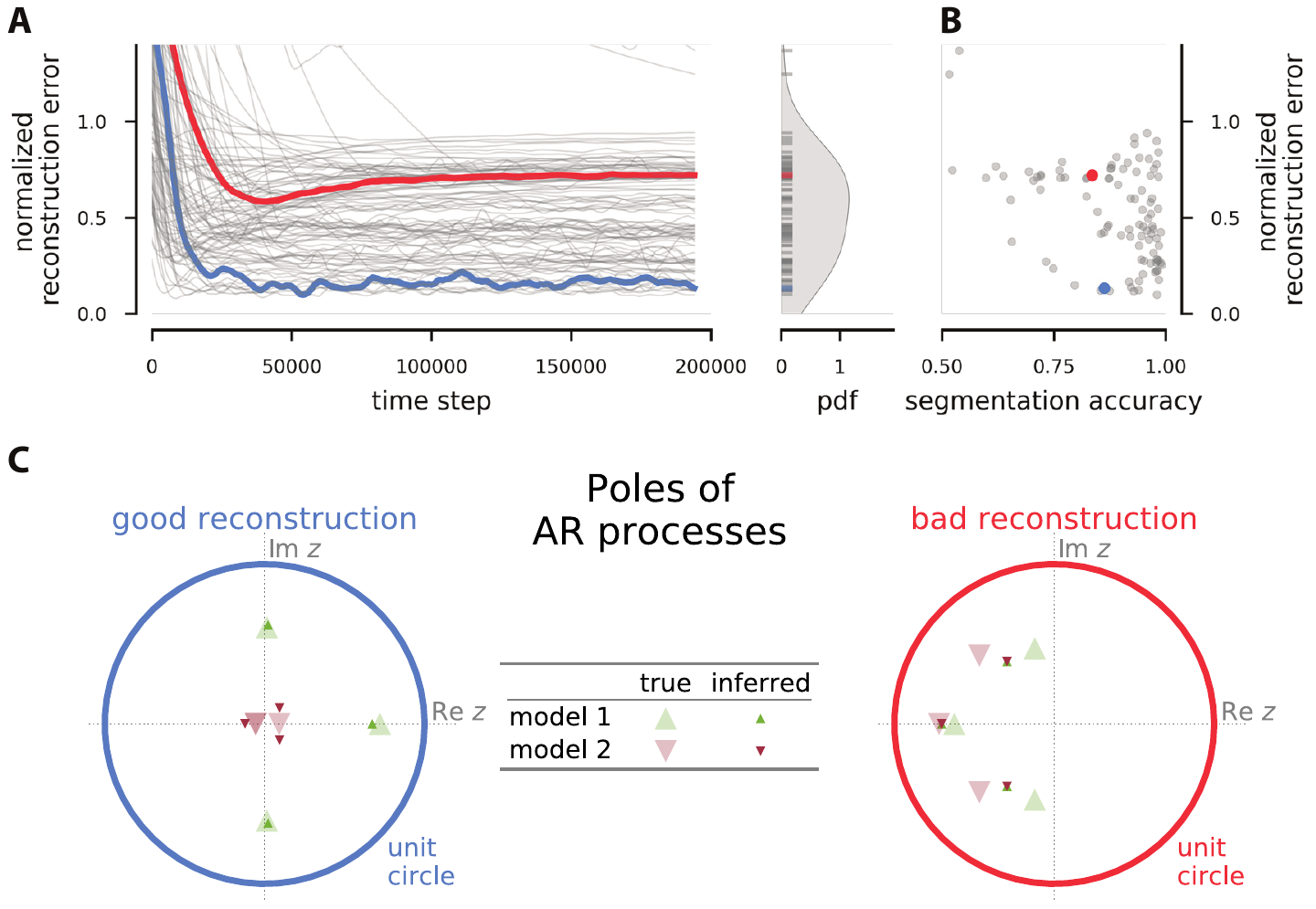}
    \caption{Weight reconstruction in the BioWTA algorithm. (A) Time evolution of the reconstruction error in the weights normalized by the difference between the ground-truth coefficients. The kernel-density plot on the right shows the distribution of final normalized reconstruction errors. (B) Relation between final normalized reconstruction error of the AR weights and segmentation accuracy. (C) Poles of the ground-truth (larger, faded triangles) and inferred (smaller, more saturated triangles) models for the runs shown in blue and red in panel A. The poles are the complex roots of the polynomial $z^p - w_1 z^{p-1} - \dotsb - w_p$ and are a convenient representation of the properties of an autoregressive model (see Appendix~\ref{sec:app:arma_details} for details). Note how in the blue run, each inferred model is close to one ground-truth model, and very different from each other. In contrast, in the red run, both inferred models are very similar to each other and interpolate between the ground-truth models.}
    \label{fig:biowta_weight_reconstruction}
\end{figure}
One of the advantages of the model-based BioWTA algorithm compared to the model-free, autocorrelation-based one is that BioWTA learns the coefficients of the generating autoregressive processes. This should in principle allow the system to predict future inputs. But how well does weight learning actually work?

In Figure~\ref{fig:biowta_weight_reconstruction}A, we show that most runs learn a noisy version of the AR weights, with deviations from the ground-truth that are smaller than the differences between the two sets of ground-truth coefficients. The accuracy of the weight reconstruction can become quite good in some cases, such as for the run highlighted in blue in the figure.

Good weight reconstruction implies high segmentation accuracy, but interestingly, the converse is often not true, as seen in Figure~\ref{fig:biowta_weight_reconstruction}B. Consider, for instance, the run highlighted in red in Figure~\ref{fig:biowta_weight_reconstruction}. The accuracy of the segmentation is as good as that for the run highlighted in blue, but its weight reconstruction is much worse.

To understand how a run can exhibit poor weight reconstruction but good segmentation accuracy, it is convenient to look at the complex roots of the characteristic polynomial $z^p - w_1 z^{p-1} - \dotsb - w_p$ where $\vect w$ are the AR coefficients for the inferred and ground-truth models. These roots are called \emph{poles} and they correspond to different modes of the dynamical system.\footnote{Note that because the characteristic polynomial is real, complex roots always appear in complex-conjugate pairs.}

Figure~\ref{fig:biowta_weight_reconstruction}C shows the poles for the inferred and ground-truth AR processes in a run with successful weight reconstruction (highlighted in blue) compared to a run where weight reconstruction failed (highlighted in red). The left panel shows good weight reconstruction: the poles for the inferred models are relatively close to the respective ground-truth values. In contrast, the right panel shows how the inferred models for the red run converged to almost a single point that interpolates between the two ground-truth models. Despite the bad weight reconstruction, the segmentation accuracy can still be high as long as the inferred models are different enough that samples from ground-truth model 1 are typically just a little bit more accurately described by inferred model 1 than inferred model 2, and \emph{vice versa}. % It would be interesting to further investigate under what circumstances this collapse of the two modes occurs and when it is drastic enough to break segmentation accuracy, but we leave these questions for future work.
%\FloatBarrier

\subsection{Some segmentation problems are intrinsically harder}
\begin{figure}[!tbh]
    \centering
    \includegraphics[width=5.76in]{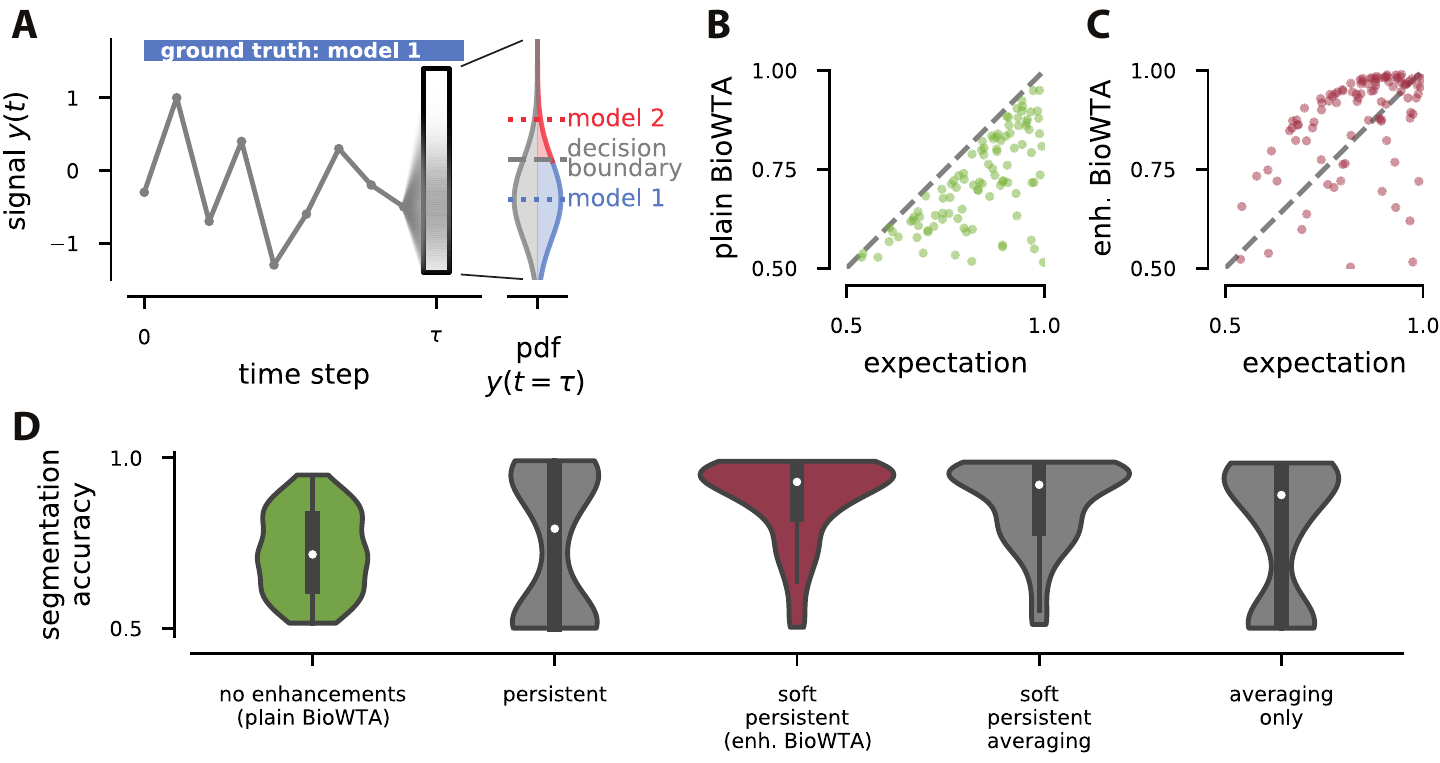}
    \caption{Understanding when and why BioWTA has trouble, and how enhancements help. (A) Sketch showing how latent-state inference can fail even if the ground-truth weights are exactly known. A fraction of samples from model 1 will predictably end up closer to the prediction from model 2 due to noise, and thus be misclassified. (B), (C) Comparison of segmentation score from our learning procedure to the naive prediction based on the argument from panel (A) (see eq.~\eqref{eq:expected_accuracy}). Panel (B) shows the results from the plain BioWTA algorithm (eq.~\eqref{eq:bio_objective_original}); panel (C) shows the improvement when using ``enhanced'' BioWTA, that is, when adding soft clustering (eq.~\eqref{eq:bio_z_update_soft}) and a persistence correction (eq.~\eqref{eq:bio_z_update_with_continuity}). (D) Kernel-density estimates of the distribution of segmentation accuracies for several variations of our algorithms. The green and dark red violins correspond to the plain and enhanced BioWTA algorithms from panels (B) and (C), respectively.}
    \label{fig:biowta_errors_and_improvements}
\end{figure}
Although the BioWTA algorithm performs very well, it is clear from Figure~\ref{fig:rolling_accuracy_biowta} that a number of runs are not so successful. In some cases, such as when the segmentation accuracy stays close to chance level, this is due to a failure in learning, which in turn can happen if, for instance, the learning rate is too large for that particular case. There is, however, significant variability even among the runs that do converge---as can be seen from Figure~\ref{fig:rolling_accuracy_biowta}C, which shows on the $x$-axis the segmentation scores of the BioWTA algorithm when the weights are kept fixed at their ground-truth values. Why is this?

The explanation for much of the variability seen in the segmentation accuracy of the BioWTA algorithm is that each randomly generated pair of AR processes can be more or less similar to each other. In the limit in which the two AR processes are identical, there would of course be no way to perform better than chance in the segmentation task. It is thus reasonable to expect that the segmentation accuracy depends on how different the two AR processes are.

This is indeed the case: even with perfect knowledge of the generating processes, segmentation will fail when a noise sample is large enough to move the signal into a range that is closer to the prediction from the wrong model. Indeed, our BioWTA algorithm infers which process the sample came from by choosing the one with the smallest prediction error.\footnote{This is also the best possible way to make the inference if the noise scales are the same for the two processes and we cannot assume anything about the relation between the states at different times.} This guess will often be correct, but it will inevitably also fail if the predictions are close enough or the noise large enough---even in the ``oracle'' case where we have perfect knowledge of the parameters describing the generating models (Figure~\ref{fig:biowta_errors_and_improvements}A).

We can derive an analytical expression for how frequently we would expect a segmentation error to occur in the ``oracle'' case, and use that to predict the segmentation accuracy. The result is the following (see Appendix~\ref{sec:app:theoretical_biowta_score}):
\begin{equation}
    \label{eq:expected_accuracy}
    \text{expected segmentation score} = \frac 12 + \frac {\arctan \bigl(\abs{\vect w_1 - \vect w_2} / \sigma \sqrt {\pi / 8}\bigr)} {\pi}\,,
\end{equation}
where $\vect w_1$ and $\vect w_2$ are the ground-truth coefficient vectors for the two processes and $\sigma$ is the standard deviation of the noise, which is chosen here such that the standard deviation of the whole signal equals 1. This guess in fact does a great job of estimating an upper bound for the segmentation accuracy of our ``plain'' BioWTA algorithm---which uses hard clustering (\emph{i.e.,} $T=0$), no persistence correction ($J = 0$), and no error averaging ($\eta_\Delta = 1$); see Figure~\ref{fig:biowta_errors_and_improvements}B.

\subsection{Algorithm enhancements boost winner-take-all performance}
Adding a persistence correction $J > 0$ significantly improves segmentation scores (see first two violins in Figure~\ref{fig:biowta_errors_and_improvements}D), at the expense of some runs failing to converge. The latter happens because the simulation can get stuck in a single state and fail to learn both generating processes. This issue can be avoided by using soft clustering instead of hard clustering (third violin, in dark red, in Figure~\ref{fig:biowta_errors_and_improvements}D). Interestingly, using soft-clustering on its own hurts rather than improve performance (see Appendix~\ref{sec:app:biowta_enhancements}). Also, adding error-averaging ($\eta_\Delta < 1$) to the soft, persistent BioWTA model can slightly hinder performance (fourth violin in Figure~\ref{fig:biowta_errors_and_improvements}D); and in fact, this ``fully-enhanced'' BioWTA model is not much better than using error-averaging on its own (last violin in \ref{fig:biowta_errors_and_improvements}D). For each variation of the algorithm, the relevant hyperparameters were optimized according to the procedure described in section~\ref{sec:biowta_is_accurate}.

\subsection{Autocorrelation-based algorithm learns faster but less accurately}
\begin{figure}[tbh]
    \centering
    \includegraphics{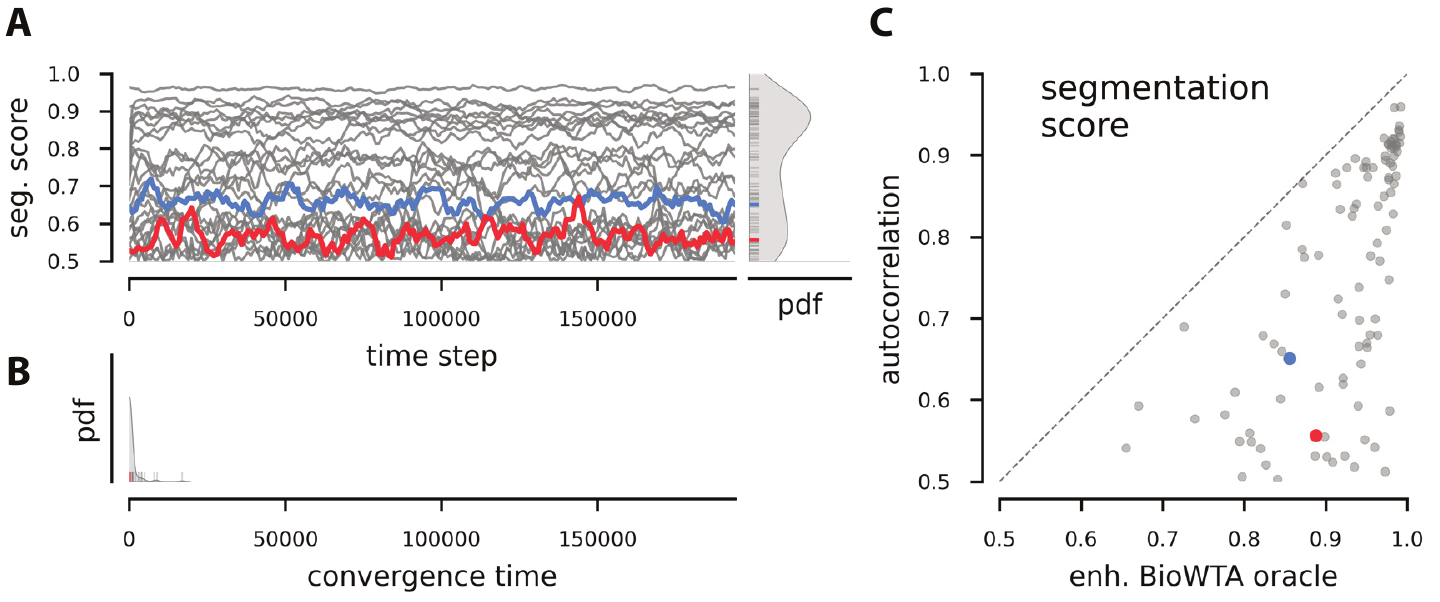}
    \caption{Segmentation accuracy for our autocorrelation-based algorithm. (A) Rolling segmentation score for a subset of 10 of the 100 simulated runs. The blue and red traces single out runs used in the other figures. A kernel-density estimate of the distribution of segmentation scores for all 100 runs is shown on the right. (B) Kernel-density estimate of the convergence times for all 100 runs. Convergence is defined as reaching a segmentation score that is at least 90\% of the final score. (C) Comparison of segmentation scores from our autocorrelation learning algorithm (on the $y$-axis) with the scores obtained from an oracle-like BioWTA algorithm where the weights are kept fixed and equal to the ground-truth values.}
    \label{fig:rolling_accuracy_xcorr}
\end{figure}
Figure~\ref{fig:rolling_accuracy_xcorr} shows the performance of the autocorrelation-based algorithm on the same 100 simulated time series used to test BioWTA above (Figure~\ref{fig:rolling_accuracy_biowta}). The segmentation is less accurate than we obtained using BioWTA, but it still reaches a median score after learning of 78\%, with 40\% of runs scoring above 85\%. Learning, however, is much faster than with BioWTA: 99\% of runs reach 90\% of their final segmentation score in less than 10,000 time steps; 84\% converge in less than 1,000 steps (Figure~\ref{fig:rolling_accuracy_xcorr}B). As for the winner-take-all algorithm, we see that generally, signals generated by pairs of less similar AR processes are easier to segment using the autocorrelation method than ones where the generating processes are very similar.

\subsection{BioWTA is competitive with oracle-like cepstral method}
\begin{figure}[tbh]
    \centering
    \includegraphics[width=5.76in]{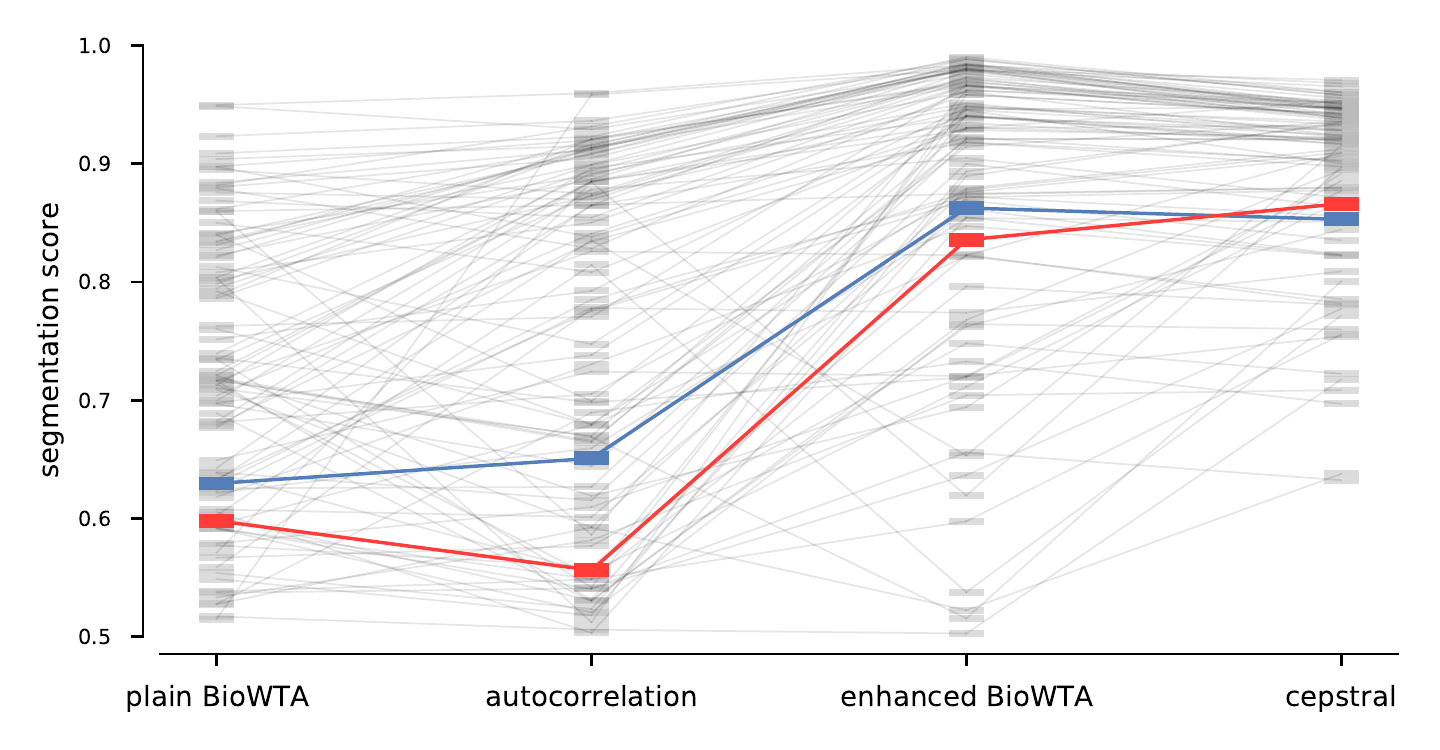}
    \caption{Comparing segmentation accuracy between our algorithms and an oracle-like cepstral method. Each thick horizontal mark indicates the segmentation accuracy estimated in the last fifth of a 200,000-sample signal. Each algorithm was tested on the same set of 100 signals drawn from alternating AR(3) processes. The thin gray lines show how the performance on the same signals compares across the algorithms. For instance, plain BioWTA and the autocorrelation method are seen to perform about the same in aggregate, but some signals are better segmented by the former while some are better segmented by the latter. In contrast, enhanced BioWTA almost always performs better than the autocorrelation method, with few exceptions. Meanwhile, the cepstral method is as accurate as enhanced BioWTA for the runs where the latter is relatively accurate, but it is not vulnerable to the cases where the coefficient learning fails, since it assumes knowledge of the ground-truth weights. Red and blue are used to highlight the same runs that were highlighted in the other figures.}
    \label{fig:algo_comparisons}
\end{figure}

Finally, we compare our algorithms with a method from control theory that uses the ground-truth weights and a cepstral measure to perform segmentation. Specifically, the inverse (moving-average) process is calculated for each ground-truth AR generating process, and the time series $y(t)$ is filtered using each inverse. When the matching inverse filter is used, the filtered output should be uncorrelated white noise. We use a cepstral norm~\citep{DeCock2002a, Boets2005} to determine how close to uncorrelated each filtered output is, and assign each time step to the model that yields the lowest cepstral norm. This method effectively relies on a rolling-window estimate for the cepstral norm, so like the autocorrelation method, it naturally takes advantage of the persistence of the latent states in our simulations. See Appendix~\ref{sec:app:cepstral_oracle} for details.

We find that our enhanced BioWTA method works basically as well as the oracle-like cepstral method, with the exception of a small fraction of runs that were likely unable to converge on a set of useful weights (see Figure~\ref{fig:algo_comparisons}). Meanwhile, the plain BioWTA and the autocorrelation-based methods work less well but can still achieve good segmentation performance on many runs.
%\FloatBarrier

\subsection{Performance is good on naturalistic stimuli}
\label{sec:vowels}
\begin{figure}[ht]
    \centering
    \includegraphics[width=5.76in]{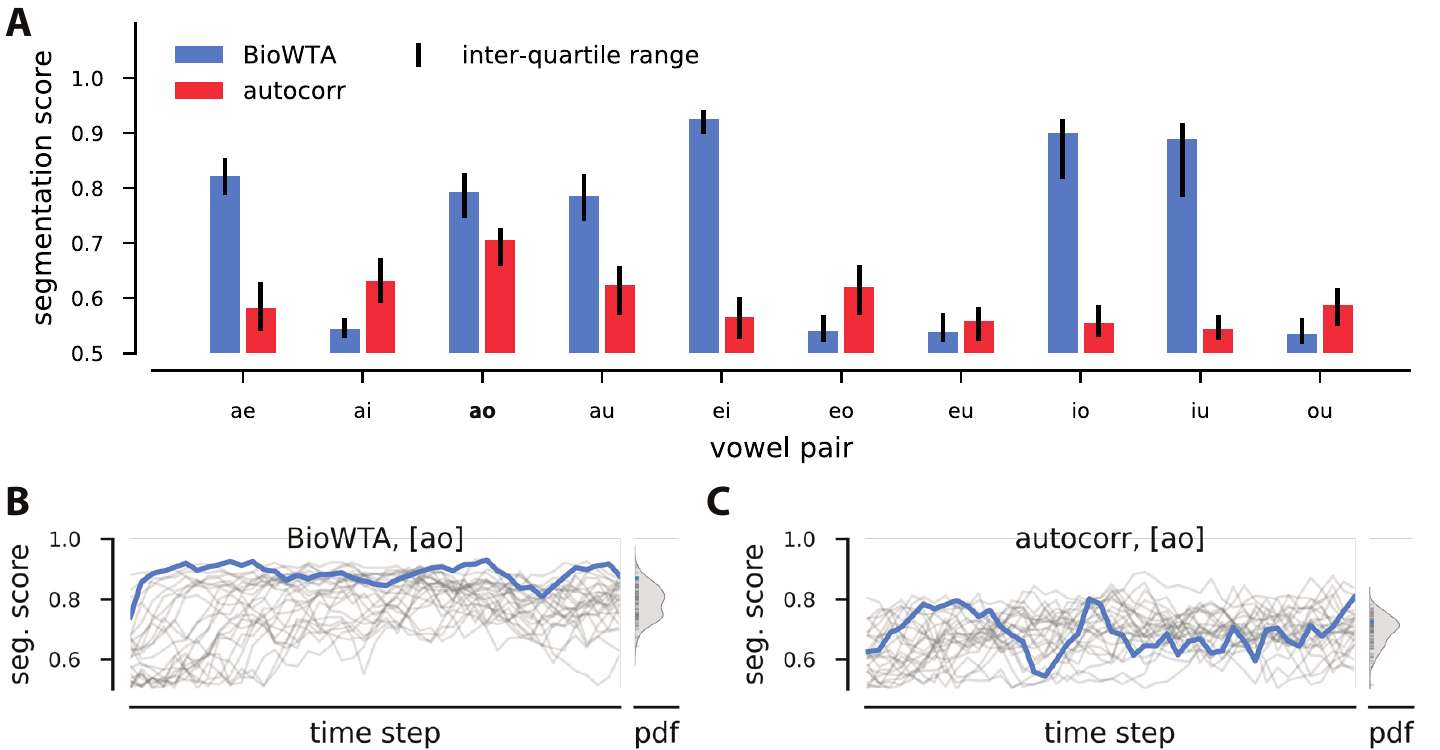}
    \caption{Segmentation accuracy of BioWTA and autocorrelation-based algorithms on datasets made by alternating snippets of voice recordings. The recordings are of vowels sung on the same note (C3); see text. (A) Median segmentation scores (colored bars) and interquartile ranges (black vertical lines) for the two algorithms on datasets obtained by alternating the 10 pairs of vowels obtained by using every combination of \textipa{[a]}, \textipa{[e]}, \textipa{[i]}, \textipa{[o]}, \textipa{[u]} (IPA notation). (B), (C) Example convergence curves and distribution of final scores for BioWTA (in (B)) and autocorrelation-based algorithm (in (C)) for the vowel pair \textipa{[ao]}.}
    \label{fig:vowel_results}
\end{figure}
To test our methods in a more realistic setting, we used datasets obtained by splicing together snippets of voice recordings. Specifically, we generated signals alternating between two different vowels sung at the same pitch (C3), with a minimum dwell time of 800 samples and an average of 1500. The voice snippets were based on recordings from \url{https://github.com/vocobox/human-voice-dataset}, downsampled to $8\,\mathrm {kHz}$ (thus, the average snippet duration amounted to about $0.2\,\mathrm{s}$). By using spliced natural signals, we obtain a more realistic test of our circuits that still gives us access to the ground truth information for the segmentation task.

Our BioWTA algorithm with two modes, $M=2$, and AR order $p=4$ reached median accuracies of around 80\% for distinguishing various pairs of vowels. Interestingly, some combinations of vowels are much harder to discriminate than others---the segmentation accuracy exceeded 90\% on the combination \textipa{[e]} and \textipa{[i]} (IPA notation), but was only slightly above chance on \textipa{[o]} \emph{vs.} \textipa{[u]} (see Figure~\ref{fig:vowel_results}A). We leave for future work a study of what properties of the vowels (\emph{e.g.,} similarity of formant frequencies) leads to these differences. Figure~\ref{fig:vowel_results}B shows convergence curves for 100 runs of BioWTA on datasets made up of alternating \textipa{[a]} and \textipa{[o]} vowels, which reach a median segmentation accuracy of almost 80\%.

The results are more nuanced for the autocorrelation algorithm. To keep the complexity of the circuit similar to that of its BioWTA counterpart, we would like to keep the order $p$ the same, $p=4$. This is similar to what we did in the previous sections. However, the meaning of the order is very different between the two algorithms, and this becomes apparent in the more realistic setting.

In the autocorrelation algorithm, $p$ sets the number of lags to use when estimating the autocorrelation. If the lags are chosen consecutively, \emph{i.e.,} $\text{lag}=1,2,\dotsc,p$, then this limits the autocorrelation length that the algorithm is sensitive to: it is at most $p$.

In contrast, the correlations in autoregressive processes can easily exceed the process order $p$: even an AR(1) can have arbitrarily long correlation lengths. For this reason, BioWTA can more easily handle setups with longer-range correlations.

We find that, indeed, when tested on vowel discrimination, the bare-bones autocorrelation-based algorithm does not fare much better than chance (median segmentation accuracy about 53\% across all vowel pairs). However, the algorithm can easily be extended to include more spacing between the lags at which the autocorrelation is estimated, \emph{i.e.,} $\text{lag} = s, 2s, \dotsc, ps$ for some constant $s$. Using relatively large steps, $s \approx 300$, allows us to increase the median segmentation accuracy to about 59\%, while keeping a low number of components, $p=4$. Specific vowel pairs can, however, be discriminated much better: the pair \textipa{[ao]} reaches segmentation accuracies of 70\%.

We speculate that the reason for which using a non-trivial step, $s > 1$, is more important here than in the AR-based datasets discussed in the previous sections is that the vowel recordings we are using are basically periodic signals with very long correlation lengths. In contrast, randomly chosen AR processes will generally have short correlation lengths. Correspondingly, the highest segmentation scores are obtained for $s=1$ when using the autocorrelation algorithm on the AR datasets.

Interestingly, the performance of the autocorrelation algorithm is in some cases complementary to that of BioWTA (see Figure~\ref{fig:vowel_results}A). For the vowel pair \textipa{[ai]}, for instance, the BioWTA segmentation accuracy is just 54\%, while the autocorrelation method reaches 63\%. This matches the results from the AR-based datasets, where the autocorrelation method was generally less accurate than BioWTA, but in some specific instances could outperform it (see Figure~\ref{fig:algo_comparisons}). It would be interesting to further investigate under what circumstances this happens, but such an analysis is beyond the scope of the current paper.

%\FloatBarrier
\section{Conclusion}

In this work we developed two biologically plausible algorithms for segmenting a one-dimensional time series based on the autoregressive processes that generated it. One algorithm is model-based and takes a normative approach, following from an optimization objective that combines clustering with model learning. This method relies on an estimate of the prediction error for both making inferences about latent states and learning the model parameters. An alternative algorithm is model-free and relies on an \emph{ad-hoc} mechanism for computing a running estimate of the autocorrelation structure of the signal. This estimate is then plugged into a clustering algorithm to achieve segmentation.

Importantly, both algorithms act online, and can be implemented in small neural networks comprised of biologically plausible units and connections with local learning rules. These provide two different architectures to look for in animal brains, depending on whether prediction error is present or not in the particular circuit under study.

Our circuits perform their task very well: the model-based, winner-take-all method achieves segmentation accuracies on-par with an oracle-like cepstral method that takes the ground-truth model parameters for granted. It also performs well when learning model parameters, although this can take many more samples than just learning to perform a good segmentation. The model-free method is less accurate than the model-based approach, but has the advantage of requiring very little training before becoming effective. This comes at the cost of not learning the model parameters in a form that can be easily used for prediction.

There are several extensions of our methods that can be readily implemented: multi-dimensional signals are a straightforward generalization; continuous signals or more complicated (but fixed) time dependencies can be handled directly by using arbitrary kernels relating the predictor vectors $\vect x$ to the signal values $y$; and signals with non-zero or even changing means can be accommodated.

It would be interesting to see to what extent our circuits can be stacked and whether this would improve their computational capabilities. One idea would be to use the error signal $\Delta$ from the BioWTA algorithm as input to another BioWTA circuit. A rather different way of using the error signal would be to map it to an action that in turn affects the dynamics of the input. We leave these ideas for future work.

Of course, nature is often not linear, so the ability to learn the parameters of non-linear dynamical systems and segment a signal based on their usage in a biologically plausible way is an important avenue for future work. Nature also does not always exhibit sharp transitions between different modalities but rather allows for gradual transitions. Adding support for such phenomena in our models would connect our work to non-negative independent component analysis (ICA), another interesting thought that we leave for future research.

\bibliography{biblio}
\bibliographystyle{apalike}

\clearpage
\numberwithin{equation}{section}
\renewcommand{\theequation}{\thesection.\arabic{equation}}
\appendix
\renewcommand{\thefigure}{S\arabic{figure}}
\setcounter{figure}{0}

\section{Derivation of the BioWTA algorithm}
\label{sec:app:biowta_derivation}

The BioWTA algorithm with all its enhancements (section \ref{sec:biowta_enhancements}) can be seen as an online approximation based on the following objective function:
\begin{equation}
    \begin{split}
        \loss &= \sum_t \sum_{k=1}^M z_k(t) \biggl\{\frac {\eta_\Delta} {2\sigma^2} \sum_{\tau=0}^t (1-\eta_\Delta)^\tau \sabs{y(t-\tau) - \vect w_k\transpose \vect x(t-\tau)}^2\\
            &\qquad - J z_k(t-1) + T\bigl(\log z_k(t) - 1\bigr)\biggr\} + \sum_t n(t)\biggl[1 - \sum_{k=1}^M z_k(t)\biggr]\\
        &= \sum_t \left\{\sum_{k=1}^M z_k(t) \biggl[\frac 1 {2\sigma^2} \Delta_k(t) - J z_k(t-1) + T\bigl(\log z_k(t) - 1\bigr) - n(t)\biggr] + n(t)\right\}\,,
    \end{split}
\end{equation}
where we added a sequence of Lagrange multipliers, $n(t)$, that help enforce the constraint $\sum_k z_k(t) = 1$ at every time step $t$. We are also assuming that the $z_k$ variables are constrained to be non-negative, $z_k(t) \ge 0$. We use the convention $0 \log 0 = 0$ to make sense of the $z_k\log z_k$ terms when $z_k = 0$.

To obtain an online algorithm, we separate the objective function into a sequence of terms, one for each time step:
\begin{equation}
    \loss = \sum_t \subloss(t)\,,
\end{equation}
with
\begin{equation}
    \subloss(t) = \sum_{k=1}^M z_k(t) \biggl[\frac 1 {2\sigma^2} \Delta_k(t) - J z_k(t-1) + T\bigl(\log z_k(t) - 1\bigr) - n(t)\biggr] + n(t)\,.
\end{equation}

We now make the online approximation by considering $\subloss(t)$ alone to be the objective function that we use when processing the $t$th sample. Differentiating with respect to $z_k(t)$ and $n(t)$ and using gradient descent-ascent yields the fast dynamics:
\begin{equation}
    \begin{split}
        \dot z_k(t) &= -\eta_z \frac {\partial \subloss(t)} {\partial z_k(t)} = -\eta_z\left[\frac 1 {2\sigma^2} \Delta_k(t) - J z_k(t-1) + T \log z_k(t) - n(t)\right]\,,\\
        \dot n(t) &= \eta_n \frac {\partial \subloss(t)} {\partial z_k(t)} = \eta_n \biggl[1 - \sum_{k=1}^M z_k(t)\biggr]\,.
    \end{split}
\end{equation}

In our simulations we do not explicitly model these fast variables, but instead directly set $z_k(t)$ to the fixed-point solution, eq.~\eqref{eq:bio_z_softmax}.

Differentiating $\subloss(t)$ with respect to the process coefficients $\vect w_k$ and using gradient descent, we get
\begin{equation}
    \label{eq:app:weight_update_general}
    \begin{split}
        \Delta \vect w_k &= -\frac {\eta_w \sigma^2} {\eta_\Delta} \frac {\partial \subloss(t)} {\partial \vect w_k} = -\frac {\eta_w} {2\eta_\Delta} z_k(t) \frac {\partial \Delta_k(t)} {\partial \vect w_k}\\
            &= -\frac {\eta_w} {2} z_k(t) \sum_{\tau=0}^t (1-\eta_\Delta)^\tau \frac {\partial} {\partial \vect w_k} \sabs{y(t-\tau) - \vect w_k\transpose \vect x(t-\tau)}^2\\
            &= \eta_w z_k(t)  \sum_{\tau=0}^t (1-\eta_\Delta)^\tau\vect x(t-\tau) \bigl(y(t-\tau) - \vect w_k\transpose \vect x(t-\tau)\bigr)\,.
    \end{split}
\end{equation}
This depends on the history of the input which we would want to avoid: in an online setting we do not want to keep many things in memory. The approach taken in the text simply ignores terms with $\tau > 0$, which makes sense if $\eta_\Delta$ is not much smaller than 1. Then eq.~\eqref{eq:app:weight_update_general} reduces to
\begin{equation}
    \Delta \vect w_k \approx \eta_w z_k(t)  \vect x(t) \bigl(y(t) - \vect w_k\transpose \vect x(t)\bigr)\,,
\end{equation}
which matches the text.

A different approach would involve keeping track of the expression
\begin{equation}
    \bar {\vect x}(t) = \sum_{\tau=0}^t (1-\eta_\Delta)^\tau\vect x(t-\tau) \bigl(y(t-\tau) - \vect w_k\transpose \vect x(t-\tau)\bigr)\,,     
\end{equation}
which is akin to an eligibility trace. This obeys
\begin{equation}
    \begin{split}
        \bar {\vect x}(t) &= \vect x(t) \bigl(y(t) - \vect w_k\transpose \vect x(t)\bigr) + \sum_{\tau=1}^{t} (1-\eta_\Delta)^\tau\vect x(t-\tau) \bigl(y(t-\tau) - \vect w_k\transpose \vect x(t-\tau)\bigr)\\
            &= \vect x(t) \bigl(y(t) - \vect w_k\transpose \vect x(t)\bigr) + \bar {\vect x}(t-1)\,.
    \end{split}
\end{equation}

Note that there is a subtlety in the expression above: in principle, the value that we use at time $t$ for $\bar {\vect x}(t')$ for $t' < t$ should depend on $\vect w_k(t)$, not on earlier values of $\vect w_k$. This would again pose problems in an online setting, so we can use the approximation
\begin{equation}
    \bar {\vect x}(t) = \vect x(t) \bigl(y(t) - \vect w_k(t)\transpose \vect x(t)\bigr) + \bar {\vect x}(t-1)\,.
\end{equation}
We do not pursue this alternative approach here.

\section{Accuracy measures}
\label{sec:app:accuracy_measures}

\paragraph{Segmentation accuracy} We define segmentation accuracy by measuring the fraction of time steps for which the inferred segmentation matches the ground-truth. We ignore the first $p$ samples for models using $p$-dimensional coefficient vectors $\vect w_k$, since no prediction can be made for an autoregressive process without having sufficient historical data.

The inferred labels can be a permutation of the ground-truth labels, since learning is unsupervised. To account for this, we use a minimum-weight matching algorithm (\verb+linear_sum_assignment+ from \verb+scipy.optimize+) to find the permutation that maximizes the segmentation score. A side effect of this is that the segmentation score cannot drop below $1 / M$, where $M$ is the number of models in the simulation.

For calculating the evolution of the accuracy score with time, we use a rolling window and apply the method described above in each window. In particular, this allows the inferred-to-ground-truth permutation to be different for different positions of the rolling window. The step by which we shift the rolling window is typically smaller than the size of the window itself. As described in the text, we use a window size of 5000 and a step of 1000 in this paper.

\paragraph{Weight-reconstruction accuracy} The weight reconstruction error is calculated by taking the differences between inferred and ground-truth coefficients, and normalizing these by the magnitude of the difference between the ground-truth values. More specifically, the error is given by:
\begin{equation}
    \text{normalized weight-reconstruction error} = \frac {\sqrt{2 \sum_k \bigl[\vect w_k^\text{inferred} - \vect w_{\sigma(k)}^\text{true}\bigr]^2}} {\bigl\lvert \vect w_2^\text{true} - \vect w_1^\text{true}\bigr\rvert}\,,
\end{equation}
where $\sigma$ is the permutation that maps each inferred model with the ground-truth model that it matches best.

The measure defined above has the useful property that if both sets of model weights converge to the same value in-between the two ground-truth coefficients, \mbox{$\vect w_k^\text{inferred}\to (\vect w_1^\text{true} + \vect w_2^\text{true}) / 2$}, then the normalized weight-reconstruction error is 1. An even larger error, $\sqrt 2$, is obtained if both inferred weights converge to a single one of the true models, $\vect w_k^\text{inferred} \to \vect w_1^\text{true}$.

The normalized weight reconstruction error is calculated using the instantaneous weight values at each time step. The error for an entire run is defined to be the error at the final time step. The time evolution of the weight reconstruction (Figure~\ref{fig:biowta_weight_reconstruction}A) employs a rolling average of the normalized reconstruction scores, with rolling-window size and step size equal to those used for calculating the rolling segmentation score (5000 steps and 1000 steps, respectively).

\section{Signal generation}
\label{sec:app:signal_generation}
We generate the signals for testing our segmentation algorithms in two steps: (1) generate the latent-state sequence, and (2) generate AR samples. The parameters of the autoregressive processes themselves are chosen randomly, as discussed below.

\paragraph{Latent-state sequence generation} We sample the latent states from a discrete-time semi-Markov model with dwell times distributed according to a truncated geometric distribution. In other words, each latent state persists for a minimum number of steps, beyond which the system switches to a different latent state with a fixed probability at each step.

\paragraph{Autoregressive sample generation} Samples are generated directly according to the model definition in eq.~\eqref{eq:regression}. For the first $p$ samples, some components of the lag vector are not defined; we define them by setting $y(t) = 0$ for $t \le 0$. Thus we can expect the first few samples of each signal to behave like a transient before stationarity is reached. After each latent-state transition, the output from the new model will depend on past samples that are generated by the old model for the first $p$ time steps.

\paragraph{Choice of autoregressive processes} We generate random autoregressive processes by starting with randomly generated poles. For this purpose, we choose a maximum pole radius $r_\text{max}$ and generate $\lfloor p / 2\rfloor$ complex numbers uniformly distributed inside the disk of radius $r_\text{max}$. These and their complex conjugates will by chosen as poles. If $n$ is odd, we additionally generate one single real pole, drawn uniformly from $[-r_\text{max}, r_\text{max}]$. We then build the monic polynomial that has these poles as roots and set it equal to $z^p - w_1 z^{p-1} - \dotsb - w_p$ to read off the coefficients $w_k$.

We typically set the maximum pole radius $r_\text{max}$ to 0.95 to ensure that the generated processes are stable.

\section{Hyperparameter optimization}
\label{sec:app:hyperopt}

\begin{figure}[!tbh]
    \centering
    \includegraphics[width=5.76in]{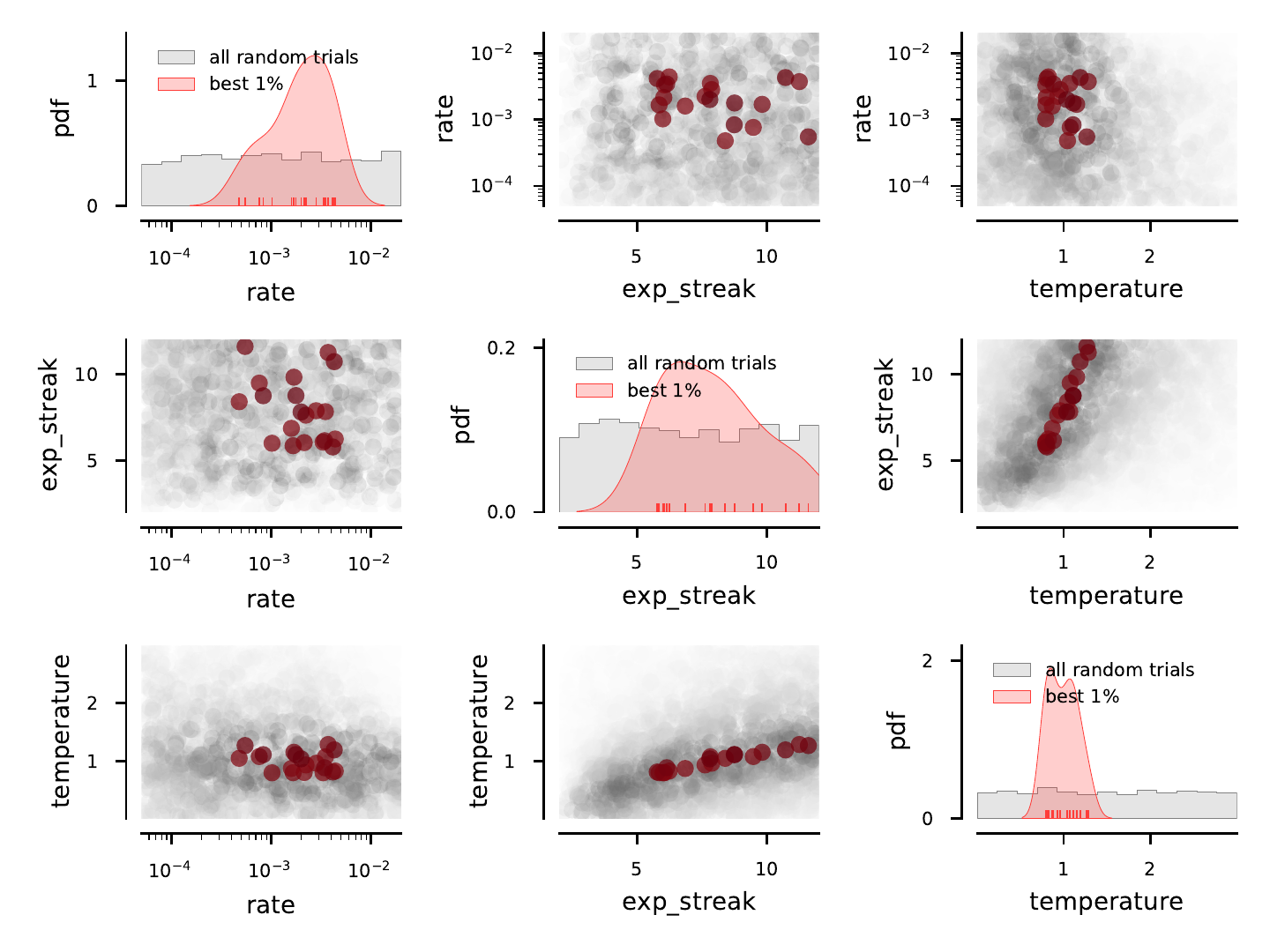}
    \caption{Hyperparameter optimization results for enhanced BioWTA on AR-based datasets. Diagonal: the distribution of sampled parameters (gray) and the distribution of those with the top fractions of successful runs (pink; see text). This shows the ranges of parameter values that lead to good segmentation. Off-diagonal: all sampled parameter values (gray) and those with the top fractions of successful runs (red; see text). This shows potential correlations between hyperparameters. The ``rate'' parameter corresponds to the learning rate $\eta_w$ from eq.~\eqref{eq:weight_updates}; the ``temperature'' is the $T$ parameter from eq.~\eqref{eq:bio_z_update_soft}; and the ``exp\_streak'' parameter is related to the persistence $J$, as described in the text.}
    % 1 - 1 / exp_streak = trans_mat = exp(J)
    \label{fig:app:enh_biowta_hyperopt_ar}
\end{figure}

The algorithms that we use depend on a number of hyperparameters, such as the learning rate $\eta_w$, the temperature $T$, and the persistence parameter $J$ in the case of enhanced BioWTA. We optimize these parameters once for every choice of algorithm and type of signal. Different choices of enhancements of BioWTA count as different algorithms.

Note that in the figures, we use a slightly different parameterization for $J$ in terms of the expected streak length ``exp\_streak'' implied by a probabilistic model whose negative log-likelihood function is minimized in eq.~\eqref{eq:wta_objective_with_continuity}. Specifically,
\begin{equation}
    \exp J = 1 - \frac 1 {\text{exp\_streak}}\,.
\end{equation}

We use two different kinds of signal: AR-based and snippet-based. The AR-based signals are generated by alternating output from several autoregressive processes, and the relevant hyperparameters are the order of the autoregressive processes and the parameters of the semi-Markov model that dictates the latent-state sequence. Throughout this paper we use only one choice for AR-based signals---AR order $p=3$, minimum dwell time 50, and average dwell time 100.

\begin{figure}[!tbh]
    \centering
    \includegraphics[width=5.76in]{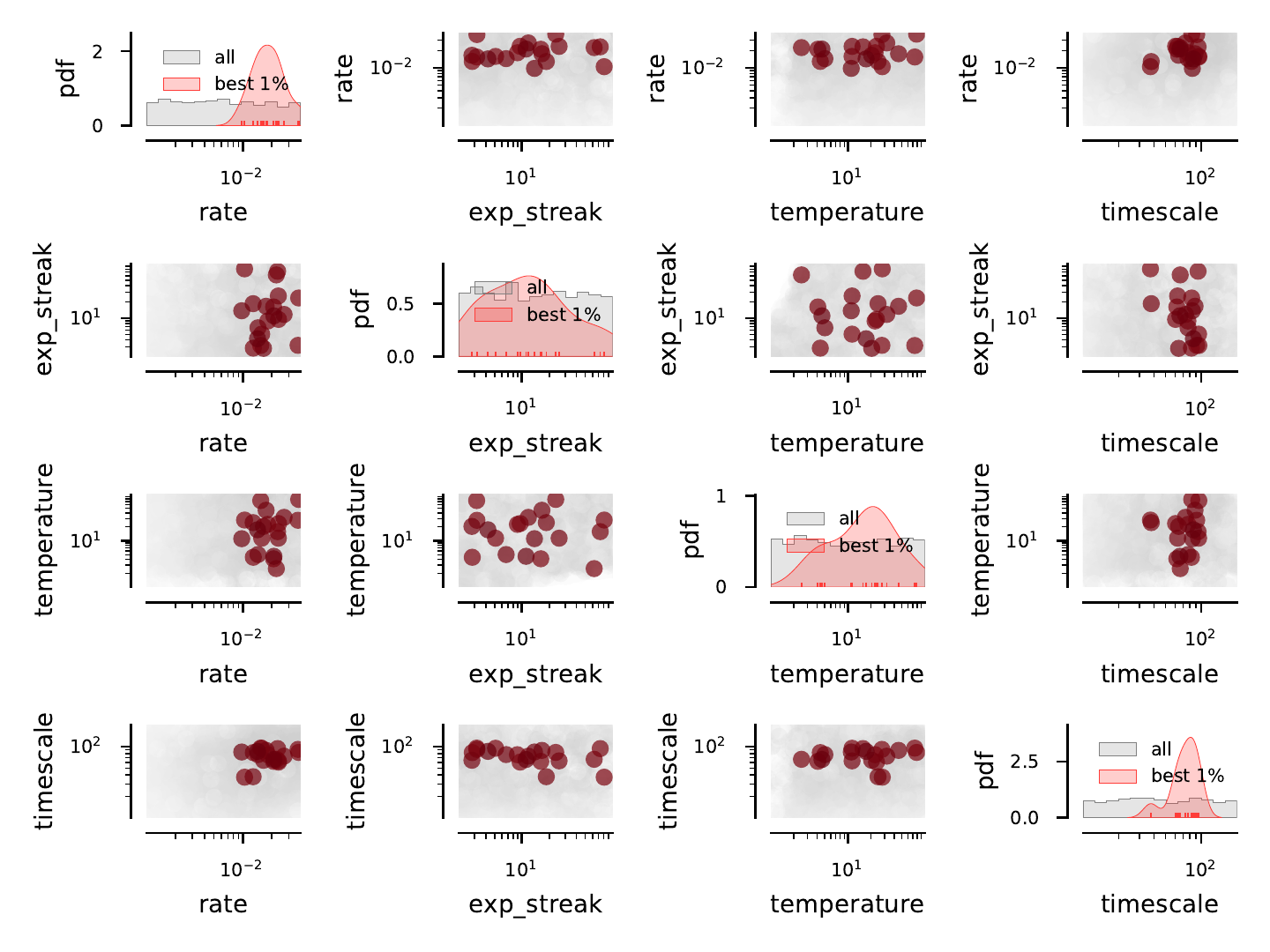}
    \caption{Hyperparameter optimization results for enhanced BioWTA on snippet-based datasets. Diagonal: the distribution of sampled parameters (gray) and the distribution of those with the top fractions of successful runs (pink; see text). This shows the ranges of parameter values that lead to good segmentation. Off-diagonal: all sampled parameter values (gray) and those with the top fractions of successful runs (red; see text). This shows potential correlations between hyperparameters. The ``rate'' parameter corresponds to the learning rate $\eta_w$ from eq.~\eqref{eq:weight_updates}; the ``temperature'' is the $T$ parameter from eq.~\eqref{eq:bio_z_update_soft}; and the ``exp\_streak'' parameter is related to the persistence $J$, as described in the text.}
    % 1 - 1 / exp_streak = trans_mat = exp(J)
    \label{fig:app:enh_biowta_hyperopt_vowel}
\end{figure}

Snippet-based signals are generated by stitching together sub-sequences cut from voice recordings of a human voice singing different vowels on a fixed pitch. The alternation of the latent states is based on a semi-Markov model, as in the AR-based datasets, but with longer minimum and average dwell times (800 and 1500, respectively).

In principle, different hyperparameters could be optimal for different pairs of vowels that are being discriminated. Instead of optimizing the parameters for each case, we run simulations on all 10 combinations of two vowels and use the median segmentation score across all these combinations for optimizing the hyperparameters.

\begin{figure}[!tbh]
    \centering
    \includegraphics[width=5.76in]{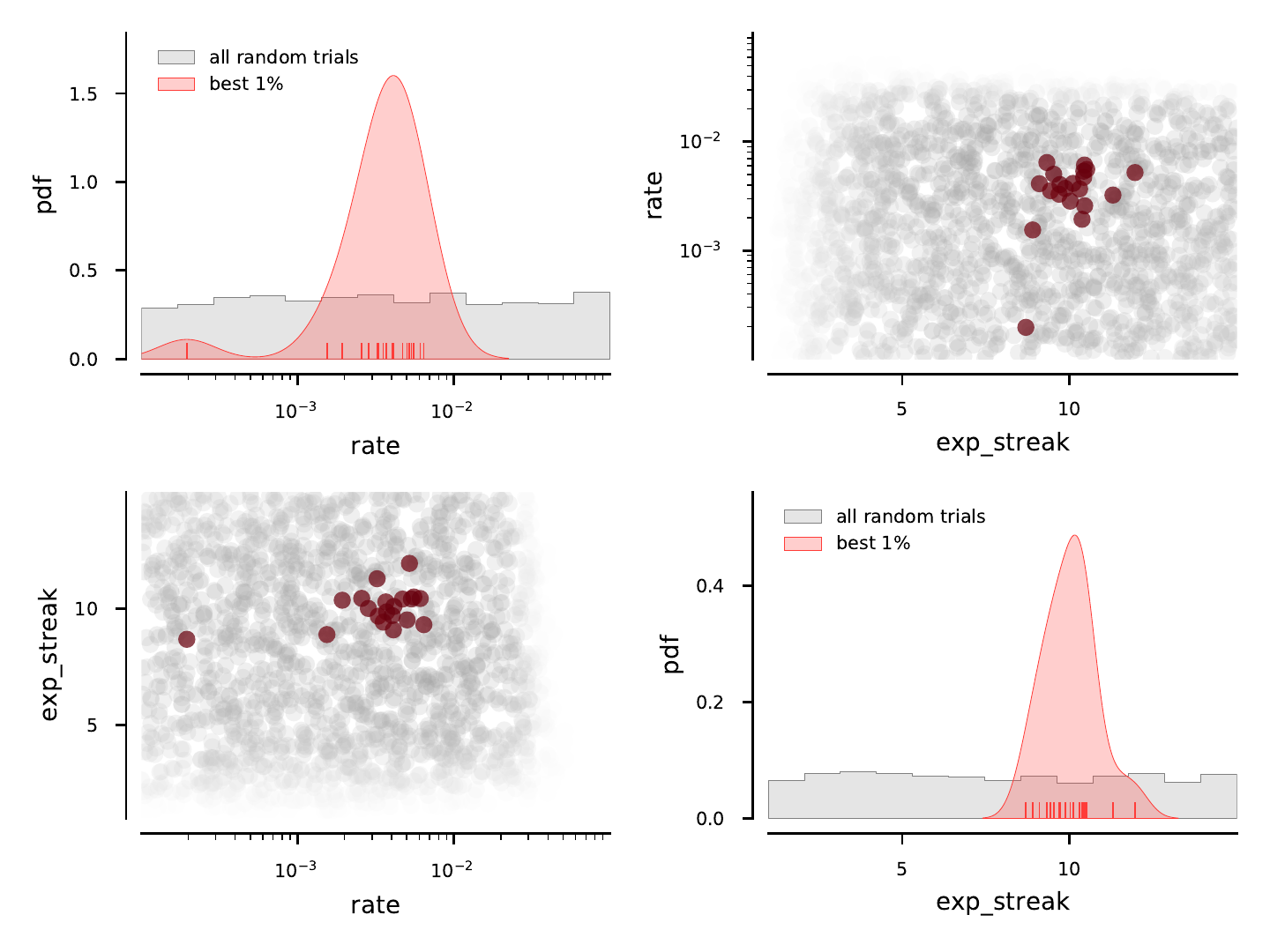}
    \caption{Hyperparameter optimization results for autocorrelation algorithm on AR-based datasets. Diagonal: the distribution of sampled parameters (gray) and the distribution of those with the top fractions of successful runs (pink; see text). This shows the ranges of parameter values that lead to good segmentation. Off-diagonal: all sampled parameter values (gray) and those with the to fractions of successful runs (red; see text). This shows potential correlations between hyperparameters. The ``rate'' parameter sets the NSM learning rate $\alpha$ in eqns.~\eqref{eq:nsm_dynamics}, while ``exp\_streak'' sets the timescale for the autocorrelation estimates, $\eta_R = \eta_\mu = 1 / \text{exp\_streak}$ (see eq.~\eqref{eq:acorrOnline}). The $\tau$ parameter from eqns.~\eqref{eq:nsm_dynamics} is fixed at $\tau = 1/2$.}
    \label{fig:app:xcorr_hyperopt_ar}
\end{figure}

We use uniform random sampling of hyperparameter values to perform hyperparameter optimization. Random search has been shown to be one of the best hyperparameter optimization methods when the number of hyperparameters is small~\citep{Bergstra2012}.

The performance of our algorithms varies due to several factors. The hyperparameter choices change the way the algorithms behave. The initial conditions affect initial transients and unlucky choices can lead to getting caught in local optima. And the signal-generation process itself is stochastic. We thus summarize the performance of the inference procedure for a batch of signals at any fixed value of the hyperparameters.

Specifically, for a number $N_h$ of randomly generated hyperparameter tuples, we choose $N_s$ random signals and measure the algorithm's segmentation accuracy on each signal at each value of the hyperparameters. We summarize the score for each hyperparameter tuple by using the fraction of ``successful'' runs, where successful is defined as having a segmentation score above some threshold $\theta_\text{good}$. This method balances the desire for runs that reach very good segmentation scores with the requirement that only a few of the runs diverge (thus receiving close to the minimum segmentation score, $1/M$).

\begin{figure}[!tbh]
    \centering
    \includegraphics[width=5.76in]{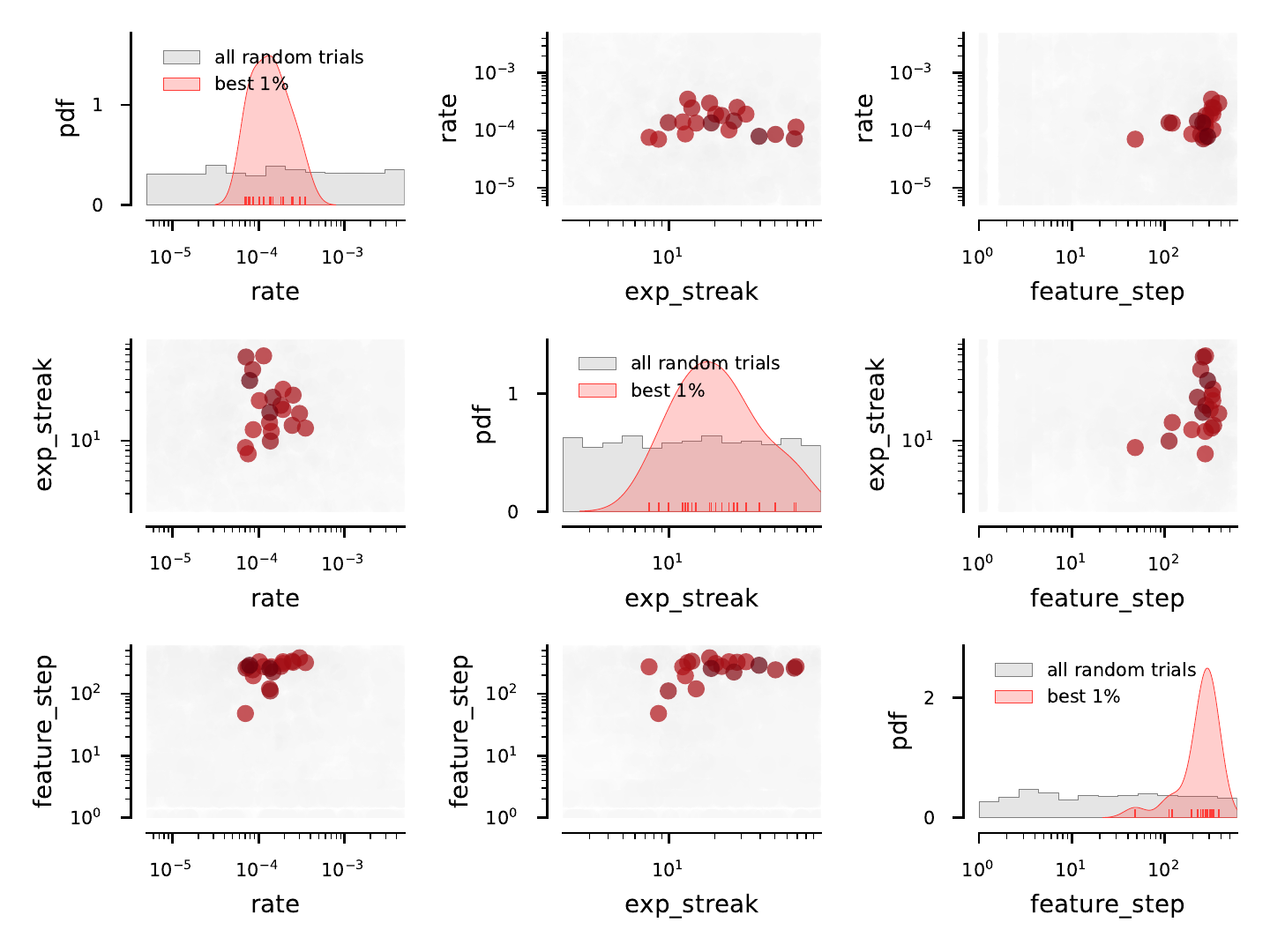}
    \caption{Hyperparameter optimization results for autocorrelation algorithm on snippet-based datasets. Diagonal: the distribution of sampled parameters (gray) and the distribution of those with the top fractions of successful runs (pink; see text). This shows the ranges of parameter values that lead to good segmentation. Off-diagonal: all sampled parameter values (gray) and those with the to fractions of successful runs (red; see text). This shows potential correlations between hyperparameters. The ``rate'' parameter sets the NSM learning rate $\alpha$ in eqns.~\eqref{eq:nsm_dynamics}, while ``exp\_streak'' sets the timescale for the autocorrelation estimates, $\eta_R = \eta_\mu = 1 / \text{exp\_streak}$ (see eq.~\eqref{eq:acorrOnline}). The $\tau$ parameter from eqns.~\eqref{eq:nsm_dynamics} is fixed at $\tau = 1/2$.}
    \label{fig:app:xcorr_hyperopt_vowel}
\end{figure}

For AR-based signals, we used $N_h = 2000$ hyperparameter choices, $N_s = 200$ signals per batch, and $\theta_\text{good} = 0.85$ to define successful runs. For the snippet-based signals, we also used $N_h = 2000$, but $N_s = 500$ signals per batch, 50 for each vowel combination. We also lowered the threshold for defining a ``successful'' run because the segmentation accuracies were generally lower; we used $\theta_\text{good} = 0.70$.

A summary of the range of tested parameters and the results of the hyperparameter optimization runs can be found in Figures~\ref{fig:app:enh_biowta_hyperopt_ar}, \ref{fig:app:enh_biowta_hyperopt_vowel}, \ref{fig:app:xcorr_hyperopt_ar}, and \ref{fig:app:xcorr_hyperopt_vowel}.

We do not attempt to make strong claims about the biological interpretation of the hyperparameter optimization in this paper. The optimization could be envisioned as the result of natural selection acting over evolutionary times; or it could be the result of a learning mechanism in the brain (\emph{e.g.,} based on reinforcement signals).

\section{Theoretical estimate of oracle BioWTA segmentation score}
\label{sec:app:theoretical_biowta_score}
To estimate how the theoretically maximum segmentation score of plain BioWTA depends on the difference between the two ground-truth AR processes, we assume an ``oracle'' setting, where the two inferred models are kept equal to their ground-truth counterparts, $\vect w_k^\text{inferred} = \vect w_k^\text{true}$, and use the argument sketched in Figure~\ref{fig:biowta_errors_and_improvements}A to estimate the fraction of misclassified samples.

More specifically, assume that a particular signal is generated by ground-truth model 1. The algorithm, however, will assign this sample to whichever process yields the lowest squared prediction error,
\begin{equation}
    z_k = \argmin_k \abs{y - \vect w_k\transpose \vect x}^2\,.
\end{equation}
Note that we are omitting the time index in this section, since it is always equal to $t$, and we are also omitting the ``true'' superscript on the ground-truth coefficients $\vect w_k$.

For model 1, the squared prediction error is simply given by the noise sample $\epsilon(t) \equiv \epsilon$ (see eq.~\eqref{eq:regression}). For model 2, we have
\begin{equation}
    \begin{split}
        \abs{y - \vect w_2\transpose \vect x}^2 &= \abs{y - \vect w_1\transpose \vect x - \Delta \vect w\transpose \vect x}^2\\
            &= \abs{\epsilon - \Delta \vect w\transpose \vect x}^2\,,
    \end{split}
\end{equation}
where
\begin{equation}
    \Delta \vect w = \vect w_2 - \vect w_1
\end{equation}
is the difference between the two ground-truth models.

Now, the sample will be assigned to the correct model (model 1) provided we have
\begin{equation}
    \epsilon^2 < (\epsilon - \Delta \vect w\transpose \vect x)^2\,.
\end{equation}
Expanding the square, this yields
\begin{equation}
    s^2 > 2 \epsilon s \,,
\end{equation}
with $s$ denoting the separation between the predictions from the two models,
\begin{equation}
    s = \Delta \vect w\transpose \vect x\,.
\end{equation}
Dividing through by $2 s$, we get
\begin{equation}
    \label{eq:app:correct_prediction_cond}
    \text{correct prediction: } \begin{cases}\epsilon < \frac 12 s & \text{if $s \ge 0$,} \\ \epsilon > -\frac 12 \abs s &  \text{if $s < 0$.}\end{cases}
\end{equation}
This corresponds to the area shaded in blue in Figure~\ref{fig:biowta_errors_and_improvements}A.

Since $\epsilon$ is drawn from a normal distribution with standard deviation $\sigma$, we can calculate the probability of a correct prediction:
\begin{equation}
    P_\text{correct} = \ncdf\left(\frac 1{2\sigma}\abs s\right) = \frac12 + \frac12 \erf\left(\frac {\abs s} {\sigma\sqrt 8}\right)\,.
\end{equation}
Note that due to the symmetry of the Gaussian distribution, this works for either sign of $s$.

The expression above tells us how accurately we can expect the cluster assignment for a given sample to be. Suppose we are now looking at an entire signal where the components of $\vect x$ are drawn from a normal distribution with standard deviation $\Sigma$. The expected accuracy is
\begin{equation}
    \label{eq:app:p_correct}
    \expect \bigl[P_\text{correct}\bigr] = \frac 12 + \frac 12 \expect\left[\erf\left(\frac {\abs {\Delta \vect w\transpose \vect x}} {\sigma \sqrt 8}\right)\right]\,.
\end{equation}
The value $\Delta \vect w\transpose \vect x$ is itself normally distributed, with mean and variance given by
\begin{equation}
    \begin{split}
        \expect\bigl[\Delta \vect w\transpose \vect x\bigr] &= 0 \,,\\
        \expect\bigl[(\Delta \vect w\transpose \vect x)^2\bigr] &= \expect\left[\sum_{i,j} \Delta w_i \Delta w_j x_i x_j\right] = \sum_{i,j} \Delta w_i \Delta w_j \text{cov}_{ij}\\
            &\sim \abs{\Delta \vect w}^2 \Sigma^2\,,
    \end{split}
\end{equation}
where in the last line we neglected cross-correlations between different components of $\vect x$. Note that these cross-correlations are not necessarily small---the fact that our signals are generated by autoregressive processes implies that these correlations are there and potentially large. Our approximation is simply meant to give a rough estimate for the expected segmentation score of plain BioWTA. Simulations suggest that our estimate is indeed quite good, Figure~\ref{fig:biowta_errors_and_improvements}B.

We thus have
\begin{equation}
    \expect \bigl[P_\text{correct}\bigr] = \frac 12 + \frac 12 \expect\left[\erf\left(\frac \Sigma \sigma \frac {\abs {\Delta\vect w}} {\sqrt 8}\,z\right)\right]\,, \quad \text{for $z\sim \mathcal {HN}(0, 1)$,}
\end{equation}
where $\mathcal{HN}(0, 1)$ is the half-normal distribution. The half-normal appears here as a direct result of eq.~\eqref{eq:app:p_correct}.

This expectation value can actually be calculated analytically (we used \texttt{Mathematica}), and the result is
\begin{equation}
    \expect \bigl[P_\text{correct}\bigr] = \frac 12 + \frac 1 \pi \arctan \left(\frac \Sigma \sigma \sqrt{\frac \pi 8}\abs{\Delta \vect w}\right)\,.
\end{equation}

Now, the expression above gives the probability of assigning a sample to, \emph{e.g.,} the first cluster \emph{provided} the sample was, in fact, generated from that cluster. In a typical simulation run used in this paper, the ground truth will alternate between the two clusters, spending about half the time in each one. This implies that the overall segmentation accuracy score should be given by
\begin{equation}
    \label{eq:app:pred_acc_exact}
    \begin{split}
        \text{predicted accuracy} &= \frac 12 + \frac 1 {2\pi} \left[\arctan \left(\frac {\Sigma_1} \sigma \sqrt{\frac \pi 8}\abs{\Delta \vect w}\right) + \arctan \left(\frac {\Sigma_2} \sigma \sqrt{\frac \pi 8}\abs{\Delta \vect w}\right)\right]\\
            &= \frac 12 + \frac 1 {2\pi} \bigl[\arctan \alpha \Sigma_1 + \arctan \alpha\Sigma_2\bigr]\,,
    \end{split}
\end{equation}
where $\Sigma_i$ is the standard deviation of the samples generated from process $i$, and we introduced the notation
\begin{equation}
    \alpha \equiv \frac 1 {\sigma} \sqrt {\frac \pi 8} \abs{\Delta \vect w}\,.
\end{equation}
In our simulations, we choose the noise standard deviation $\sigma$ such that the overall variance of the output is 1. Since the generating process is split 50-50 between the two possible latent states, this implies
\begin{equation}
    \label{eq:app:var_sum}
    \frac 12 \bigl(\Sigma_1^2 + \Sigma_2^2\bigr) = 1\,.
\end{equation}
The specific values for $\Sigma_i$ will depend on the run, but in order to get a rough answer (that will turn out to work well in practice), we choose the case in which the two are approximately equal,
\begin{equation}
    \Sigma_1 \approx \Sigma_2 = 1\,.
\end{equation}
This means that we can estimate
\begin{equation}
    \text{predicted accuracy} \approx \frac 12 + \frac 1 {\pi} \arctan \alpha \equiv \frac 12 + \frac 1 {\pi} \arctan \frac 1 {\sigma} \sqrt {\frac \pi 8} \abs{\Delta \vect w}\,.
\end{equation}
This is the formula we used in the paper.

\paragraph{General case} Employing a standard trigonometric identity and using the fact that $\alpha$, $\Sigma_1$, and $\Sigma_2$ are all positive, we can rewrite eq.~\eqref{eq:app:pred_acc_exact} as\footnote{We use the $\arccot$ here instead of $\arctan$ because the range of $\arctan$ is $[-\pi/2, \pi/2]$, while the sum $\arctan \alpha\Sigma_1 + \arctan \alpha\Sigma_2$ can range from 0 to $\pi$.}
\begin{equation}
    \text{predicted accuracy} = \frac 12 + \frac 1 {2\pi} \arccot \frac {1 - \alpha^2 \Sigma_1 \Sigma_2} {\Sigma_1 + \Sigma_2}\,.
\end{equation}
From eq.~\eqref{eq:app:var_sum}, we find
\begin{equation}
    \label{eq:app:std_sum}
    (\Sigma_1 + \Sigma_2)^2 = 2(1 + \Sigma_1 \Sigma_2)\,,
\end{equation}
which, using the notation
\begin{equation}
    \theta \equiv \frac {\Sigma_1 + \Sigma_2} 2 \,,
\end{equation}
yields
\begin{equation}
    \text{predicted accuracy} = \frac 12 + \frac 1 {2\pi} \arccot \frac {1 - \alpha^2 (2\theta^2 - 1)} {2 \alpha \theta}\,.
\end{equation}
Now, eq.~\eqref{eq:app:std_sum} shows that the sum of the two standard deviations, $\Sigma_1 + \Sigma_2$, is at least $\sqrt 2$. Put differently, $\theta \ge \frac {\sqrt 2} 2\approx 0.71$. Conversely, the product of two numbers with a fixed sum is maximum when the two numbers are equal to each other, which, in combination with eq.~\eqref{eq:app:var_sum}, implies that $\Sigma_1^2 \Sigma_2^2 \le 1$. This in turn means that $\theta \le 1$. All in all, the mean standard deviation is confined to a rather small range of values:
\begin{equation}
    \frac {\sqrt 2} 2 \le \theta \le 1\,.
\end{equation}

\section{Performance effects of BioWTA enhancements}
\label{sec:app:biowta_enhancements}
\begin{figure}[!tbh]
    \centering
    \includegraphics[width=5.76in]{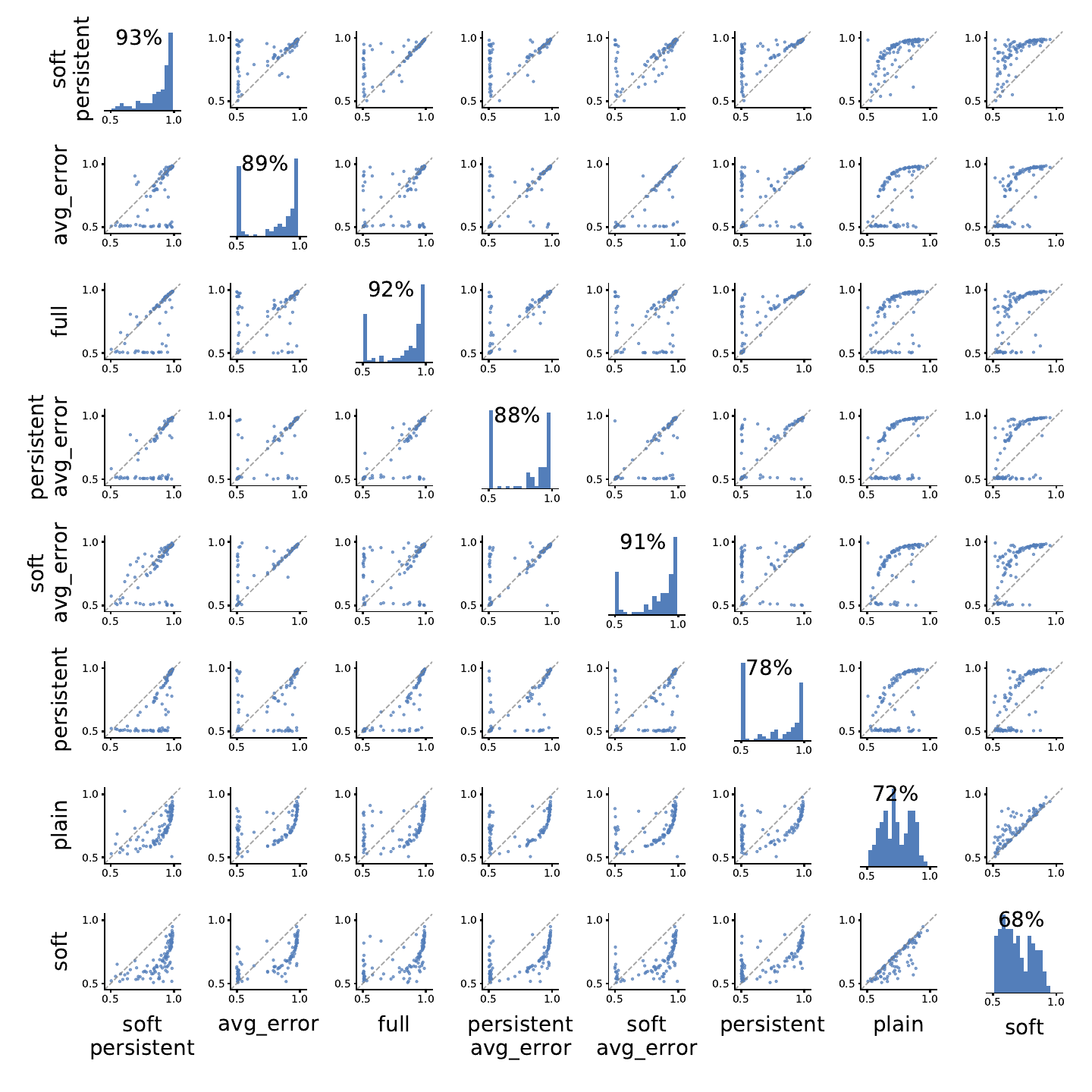}
    \caption{Comparison of segmentation accuracy for all combinations of BioWTA enhancements. The plot at position $(i, j)$, for $i \ne j$ in the figure compares the segmentation accuracy from method $i$ (on the $y$-axis) to that from method $j$ (on the $x$-axis). The plots on the diagonal (\emph{i.e.,} when $i = j$) are histograms showing the distribution of the accuracy scores for each method. The number above each histogram is the median segmentation score obtained for that method.}
    \label{fig:app:biowta_mods_comparison}
\end{figure}

We saw in the text that the enhancements we described for the BioWTA method---the persistence correction controlled by the $J$ parameter; the soft clustering controlled by the temperature $T$; and the averaging of the squared error controlled by the $\eta_\Delta$ parameter---affect performance in non-trivial way. Figure~\ref{fig:app:biowta_mods_comparison} shows this in more detail.

In particular, note that soft clustering on its own has a consistent detrimental effect on segmentation performance, but the highest-scoring method includes soft clustering in addition to the persistence correction. This latter method performs significantly better than if the persistence correction is used on its own (see first, sixth, and last row and column in Figure~\ref{fig:app:biowta_mods_comparison}).

The error-averaging correction has the opposite behavior: on its own it yields results almost as good as the top-performing method, although it is more vulnerable to convergence failure. On the other hand, when combined with the other enhancements, it fails to improve the high-performing runs and instead hurts performance by hindering convergence in some runs (see first through fourth row and column in Figure~\ref{fig:app:biowta_mods_comparison}). The only exception is when used in conjunction with soft clustering, which works better than either having only one of the enhancements, or none at all (see second, fifth, and last row and column in Figure~\ref{fig:app:biowta_mods_comparison}).

Persistence on its own improves performance significantly for many runs, but leads to convergence problems in other runs (see sixth and seventh row and column in Figure~\ref{fig:app:biowta_mods_comparison}). It behaves almost identically to the error-averaging correction alone when used in conjunction with it (second and fourth row and column in Figure~\ref{fig:app:biowta_mods_comparison}), but it has the best performance of all the methods we've tried when used with soft clustering (first and last row and column in Figure~\ref{fig:app:biowta_mods_comparison}).

\section{Some details about ARMA processes}
\label{sec:app:arma_details}

\paragraph{ARMA processes and inverses} It can be useful to think of an extension of AR models, the autoregressive moving-average (ARMA) process. These involve  a weighted moving average (MA) of the noise signal in addition to the autoregressive part,
\begin{equation}
    \label{eq:app:arma}
    y(t) =  w_1 y(t-1) + \dotsb + w_p y(t-p) + \epsilon(t) + b_1 \epsilon(t-1) + \dotsb + b_q \epsilon(t-q)\,.
\end{equation}

The output of an AR process can be inverted using an MA process to get back at the noise sequence $\epsilon(t)$:
\begin{equation}
    \label{eq:app:ar_inverse}
    \begin{split}
        \text{if } y(t) &= w_1 y(t-1) + \dotsb + w_p y(t-p) + \epsilon(t)\,,\\
        \text{then } \epsilon(t) &= y(t) - w_1 y(t-1) - \dotsb - w_p y(t-p)\,.
    \end{split}
\end{equation}
More generally, any ARMA process admits an inverse (though the inverse process might be unstable). This is relevant for the cepstral oracle method described in the next section.

\paragraph{Spectral properties: poles and zeros} The signal $y(t)$ and noise $\epsilon(t)$ enter linearly in the definition of an ARMA process, eq.~\eqref{eq:app:arma}, with various delays. Because of this, a $z$-transform is useful for analyzing ARMA processes:
\begin{equation}
    Y(z) = \sum_t y(t) z^{-t}\,,
\end{equation}
where $z$ is a complex number. This can be related to the Fourier series (or frequency-space representation) of the signal by focusing on the unit circle, $z = e^{-2\pi i f}$.

The transformation induced by an ARMA process has a simple form after a $z$-transform:
\begin{equation}
    Y(z) = H(z) E(z)\,,
\end{equation}
where the \emph{transfer function} $H(z)$ is given by
\begin{equation}
    H(z) = \frac {1 + b_1 z^{-1} + \dotsb  b_q z^{-q}} {1 - w_1 z^{-1} - \dotsb w_p z^{-p}} \equiv \frac {B(z)} {W(z)}\,.
\end{equation}

The transfer function blows up at roots of the denominator $W(z)$, which is why they are also called the ``poles'' of the system. The magnitude of a pole is related to the temporal extent of the response to a particular excitation; poles outside the unit circle give rise to instabilities in the ARMA process.

The transfer function vanishes at the roots of the numerator $B(z)$, so these are called the ``zeros'' of the system. Inverting an ARMA process swaps the poles with the zeros, so a system with a stable inverse must have all the zeros contained within the unit circle.

\section{Segmentation asymmetries with BioWTA}
\label{sec:app:asymmetry}
\begin{figure}[!ht]
    \centering
    \includegraphics[width=5.76in]{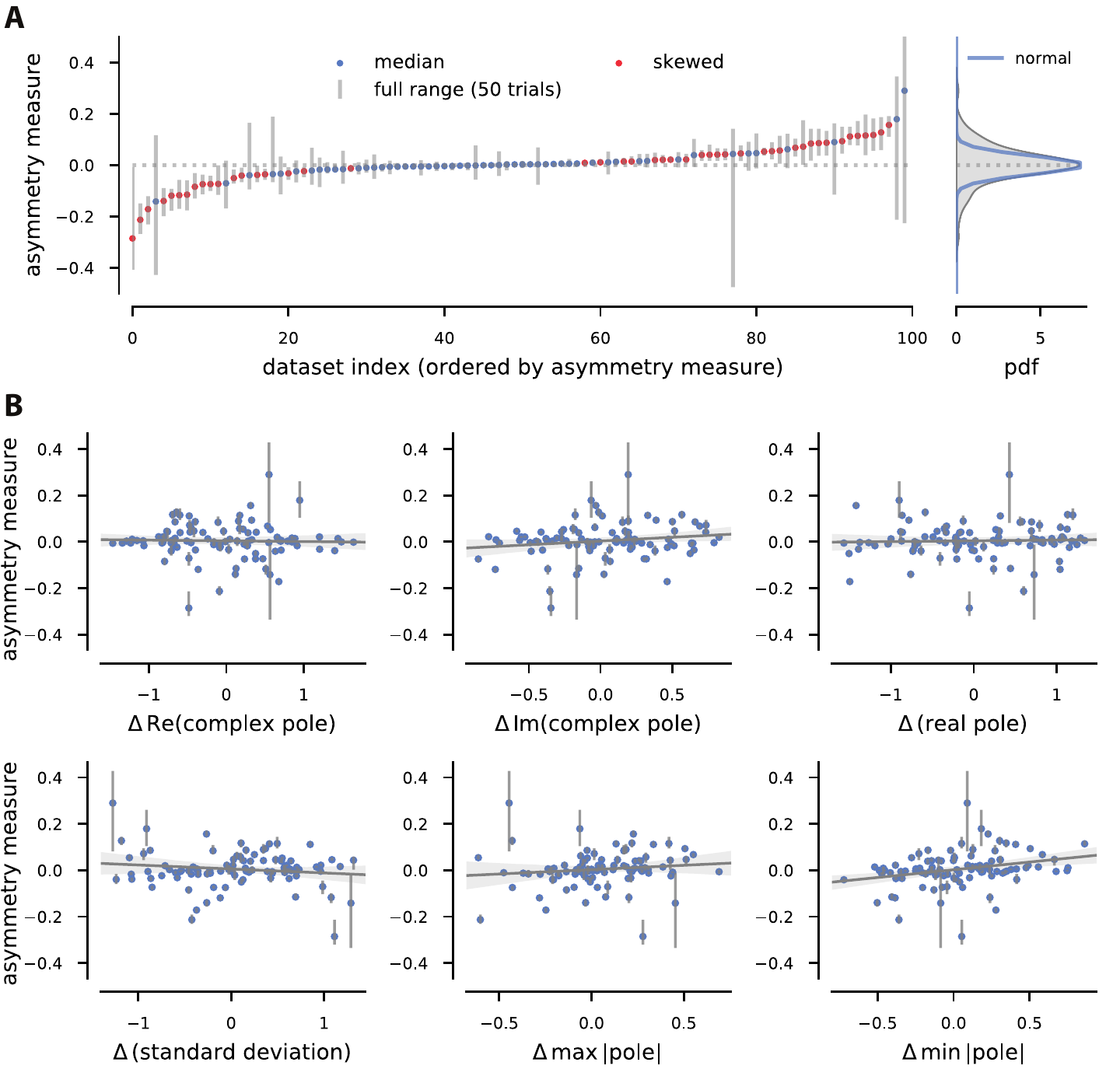}
    \caption{Asymmetric BioWTA segmentation performance. (A) Asymmetry score (difference between fraction of time when each process is mislabeled) for 100 datasets (50 runs each), ordered by median asymmetry. Colored dots: median asymmetry; gray lines: range of asymmetry scores per dataset. Red: datasets where all asymmetry scores have the same sign. Right panel: distribution of asymmetry scores, with zero-centered Gaussian matching central peak (blue). (B) Scatter plots showing how the asymmetry measure correlates with some characteristics of the AR processes being discriminated. See text below for details.}
    \label{fig:asymmetry_results}
\end{figure}
The segmentation results generated by our methods can exhibit a certain level of asymmetry: one of the two processes is more often mislabeled than the other one. To quantify this effect, we start by calculating the fraction of time steps where each ground-truth process is mislabeled.\footnote{We match ground-truth to predicted labels by using the assignment that maximizes overall segmentation performance.} We then take the difference between the two fractions as a measure of the asymmetry in the segmentation performance. To avoid transient effects due to learning, we use only the last 10\% of samples in each run to estimate the asymmetry.

We might expect some asymmetry to result from stochastic effects related to the initialization of the model or the specific choice of latent-state sequence. We therefore generate 50 different signals for each choice of pair of AR(3) models and run BioWTA on each of them. We do this for 100 randomly chosen pairs of AR(3) processes.

Interestingly, there appear to be asymmetries in segmentation performance that go well beyond stochastic effects: roughly half of the 100 pairs of processes result in asymmetry measures that are either all positive or all negative in the entire set of 50 signals that we generated for each pair (see Figure~\ref{fig:asymmetry_results}A). It thus seems that in some combinations, some processes are harder to identify than others.

It appears difficult to understand exactly what aspect of the AR process makes the difference. In Figure~\ref{fig:asymmetry_results}B we attempted to predict the asymmetry score (on the $y$-axis) using a variety of measures: the real and imaginary parts of the complex pole\footnote{Since the AR processes we're using have order $p=3$ and real coefficients, there are (almost surely) two complex poles that are conjugates of each other, and one real pole. Of the pair of complex poles, we choose the one with the positive imaginary part.}, the real pole, the standard deviation of the signal generated by each process, and the minimum and maximum pole magnitude.  More specifically, we used the difference between each such measure for each process. Of all the measures we used, only the standard deviation and the minimum pole magnitude appear to have statistically significant predictive power at the 0.05 level, and they are not particularly good predictors of asymmetry: together, all the factors we considered account for only about 15\% of the total variance in the asymmetry measure.

It would be interesting to further investigate the characteristics of pairs of AR processes that can explain which one is easier to identify in a segmentation task, but this is beyond the scope of the current paper. We leave this question for future work.

\section{Cepstral oracle method}
\label{sec:app:cepstral_oracle}

\paragraph{General overview} A standard control-theory method for fault detection relies on using the inverse of a system with a known transfer function to detect anomalies in the functioning of the system~\citep{DeCock2002a, Boets2005}.

Specifically for our purposes, consider the signal from eq.~\eqref{eq:regression},
\begin{equation}
    y(t) = \sum_{k=1}^M z_k(t) \Bigl[\vect w_k\transpose \vect x(t) + \epsilon(t)\Bigr]\,.
\end{equation}
Each of the AR processes defined by the coefficients $\vect w_k$ has an inverse MA process with coefficients $\vect b_k = -\vect w_k$, as shown in eq.~\eqref{eq:app:ar_inverse}. We can filter the signal $y(t)$ using each of these inverse filters to obtain
\begin{equation}
    \label{eq:app:filterings}
    \epsilon_k(t) = y(t) - \vect w_k\transpose \vect x(t)\,.
\end{equation}
Notice that this is nothing else but the prediction error in the ``oracle'' setting where the model weights $\vect w_k$ are set to their ground-truth values.

If the actual process that generated the sample at time $t$ is $\hat z(t)$, then $\epsilon_{\hat z(t)} \equiv \epsilon(t)$, \emph{i.e.,} uncorrelated Gaussian noise. In contrast, all other filterings, $\epsilon_k(t)$ for $k\ne \hat z(t)$, will still contain temporal correlations.

The cepstral oracle method relies on a measure of the strength of temporal correlations to find the index $k$ that leads to the least temporally correlated filtering $\epsilon_k(t)$. This provides a best guess for the identity of the generating process.

\paragraph{Cepstral norm} The specific measure of temporal correlation that we use here is a \emph{cepstral norm}~\citep{DeCock2002, DeCock2002a, Boets2005}.

The (power) cepstrum is the inverse Fourier transform of the logarithm of the power spectrum of a signal~\citep{DeCock2002a},
\begin{equation}
    \label{eq:app:cepstrum}
    c(k) = \sabs{\mathcal F^{-1} \Bigl(\log \sabs{\mathcal F(y(t))}^2\Bigr)}\,,
\end{equation}
where $\mathcal F$ denotes the Fourier transform operator. This has the convenient property that it turns convolutions into sums, thus allowing to separate different stages of filtering if these have different-enough spectral responses.

If we define the \emph{cepstral norm}
\begin{equation}
    \label{eq:app:cepstral_norm}
    g(y)^2 = \sum_{k=0}^\infty k \abs{c(k)}^2\,,
\end{equation}
this provides a measure of the distance between the signal $y(t)$ and uncorrelated Gaussian noise~\citep{DeCock2002a, Boets2005}. In practice, only a finite number of cepstral coefficients are used in the calculation.

\paragraph{Calculating the cepstral norm} A sequence of samples from the signal $y(t)$ are necessary for calculating the cepstral norm. Applying the definitions~\eqref{eq:app:cepstrum} and~\eqref{eq:app:cepstral_norm} directly does not lead to the most efficient estimate. Instead, we start by defining ``past'' and ``future'' \emph{Hankel matrices},
\begin{equation}
    \label{eq:app:past_future_hankel}
    \begin{split}
        Y_p &= \frac 1 {\sqrt \tau} \begin{pmatrix}y(0) & y(1) & \hdots & y(\tau-1) \\
			    y(1) & y(2) & \hdots & y(\tau) \\
			    \vdots & \vdots & \ddots & \vdots \\
			    y(k-1) & y(k) & \hdots & y(k + \tau - 2)\end{pmatrix}\,, \\
	    Y_f &= \frac 1 {\sqrt \tau} \begin{pmatrix}y(k) & y(k + 1) & \hdots & y(k + \tau-1) \\
			    y(k + 1) & y(k + 2) & \hdots & y(k + \tau) \\
			    \vdots & \vdots & \ddots & \vdots \\
			    y(2k-1) & y(2k) & \hdots & y(2k + \tau - 2)\end{pmatrix}\,,
	\end{split}
\end{equation}
where $k$ gives the maximal cepstral order used in the cepstral norm formula, eq.~\eqref{eq:app:cepstral_norm}, and $\tau$ gives the number of samples over which we're averaging. We also define a ``total'' Hankel matrix
\begin{equation}
    \label{eq:total_hankel}
    Y =  \begin{pmatrix}Y_p \\ \hdashline Y_f \end{pmatrix}\,.
\end{equation}
In terms of the Hankel matrices, the cepstral norm can be approximated by
\begin{equation}
    \label{eq:app:cepstral_norm_as_logdet}
    g(y)^2 \approx \log \det Y_p Y_p^T + \log \det Y_f Y_f^T - \log \det Y Y^T\,,
\end{equation}
which becomes exact in the limit $k\to\infty$, $\tau\to\infty$. Furthermore, the calculation of the determinants can be simplified by using LQ decompositions,
\begin{equation}
    \label{eq:app:hankel_lq}
    Y_p = L_p Q_p \,, \qquad Y_f = L_f Q_f \,, \qquad Y = L Q\,,
\end{equation}
where $L$ are lower-triangular matrices and $Q$ are orthogonal matrices. With these notations, we can write
\begin{equation}
    \label{eq:app:cepstral_norm_from_lq}
    \begin{split}
        g(y)^2 &\approx 2 \sum_{j=1}^k \bigl(\log L_{jj}^p + \log L_{jj}^f\bigr) - 2 \sum_{j=1}^k \log L_{jj}\\
            &= 2 \sum_{j=1}^k \log L_{jj}^f - 2 \sum_{j=k+1}^{2k} \log L_{jj}\,,
    \end{split}
\end{equation}

\paragraph{Rolling estimate of the cepstral norm} In our setting, the generating process changes during the duration of the signal. To find a local estimate of how uncorrelated a filtering $\epsilon_k(t)$ from eq.~\eqref{eq:app:filterings} is, we could apply the cepstral norm calculation in sliding window, much like we do when we calculate the segmentation score. A more efficient approach uses a discounting mechanisms akin to an exponential moving average, and can be implemented online. We will not give all the details here, but it relies on a redefinition of the Hankel matrices:
\begin{equation}
    \tilde Y_{ij}(t) = \gamma^{\tau - j} y(t + i + j) = \gamma^{\tau-j} Y_{ij}(t)\,,
\end{equation}
where the indices are assumed to be zero-based. This discounts older samples by a factor of $\gamma$ raised to the number of time steps that have passed since those samples were observed.

Whenever a new sample is obtained, a column is appended to the Hankel matrix, and the rest of the elements are discounted by an additional factor of $\gamma$,
\begin{equation}
    \tilde Y_{ij}(t+1) = \gamma^{\tau - j} y(t + i + j + 1) = \frac {\gamma^{\tau-j}} {\gamma^{\tau - j - 1}} \tilde Y_{i,j + 1}(t) = \gamma \tilde Y_{i, j + 1}(t)\,.
\end{equation}
Since for the cepstral norm calculation we are only interested in the $L$ factor of the LQ decomposition of $Y$ (as in eq.~\eqref{eq:app:cepstral_norm_from_lq}), we can actually use an algorithm for updating the LQ decomposition based on Givens rotations~\citep{oppenheim2001discrete} to calculate the effect of appending a column. The effect of multiplying the Hankel matrix by $\gamma$ is simply to multiply $L$ by the same factor.

Details of these procedure can be found in the implementation available on GitHub, at \url{https://github.com/ttesileanu/bio-time-series}.

\section{Straightforward generalizations of BioWTA}
\label{sec:app:generalizations}
\paragraph{Processes with non-zero mean}
One way to lift the assumption on the mean is by using an adaptation mechanism to subtract the mean from the data before segmentation. The rest of our algorithm would stay the same.

A different approach would be to add an extra dimension to the input $\vect x$, a component that is constant in time. This would allow the corresponding component of the weight vector $\vect w$ to encode a bias proportional to the process mean, without further changes to our method.

\paragraph{More complex history dependencies}
We can use arbitrary, though fixed, functions of the past in the model, \emph{i.e.,}
\begin{equation}
    \begin{split}
        y(t) &= \sum_{k=1}^M z_k(t) \Bigl[\vect w_k\transpose \vect x(t) + \epsilon(t)\Bigr]\,,\\
        x_i(t) &= g_i(y(t-1), y(t-2), \dotsc)\,,
    \end{split}
\end{equation}
where $g_i$ can be general functions. These can be implemented by neural circuits that are evolutionarily encoded. Alternatively, they could be learned by a different mechanism (\emph{e.g.,} reinforcement learning) on longer timescales. The model from the main text maps to $g_i(t) = y(t - i)$.

A similar procedure allows us to extend our methods to continuous time: we would use a set of fixed kernels $K_i$ and define
\begin{equation}
    x_i(t) = K_i(t) * y(t)\,,
\end{equation}
where $*$ denotes the convolution operator. Arbitrary kernels can be implemented by stacking leaky integrator neurons with appropriate decay times.

Note that both of the approaches described above work without further modifications to the algorithm.

\paragraph{Multidimensional signals}

The generalization to multidimensional signals is in principle straightforward: simply promote the output $y(t)$ to a vector, turning the weights $\vect w$ into matrices and the noise term $\epsilon$ into a vector:
\begin{equation}
    \vect y(t) = \sum_{k=1}^M z_k(t) \Bigl[\mathbf W_k\transpose \vect x(t) + \vect \epsilon(t)\Bigr]\,.\\
\end{equation}
The learning rules generalize without complication:
\begin{equation}
    \begin{split}
        z_k(t) &= \softmax{}_T \left\{-\frac 1 {2\sigma^2} \bigl\lvert \vect y(t) - \mathbf W_k(t)\transpose \vect x(t)\bigr\rvert^2\right\}\,,\\
        \mathbf W_k(t+1) &= \mathbf W_k(t) + \eta_w z_k(t) \vect x(t) \vect \Delta_k(t)\transpose\,,\\
        \vect \Delta_k (t) &= \vect y(t)  - \mathbf W_k(t)\transpose \vect x(t)\,.
    \end{split}
\end{equation}
Note that the locality of the synaptic plasticity rule is preserved.

The number of parameters involved in such a model can quickly grow to unmanageable levels. To avoid this, we can impose further structure on the weight matrices, such as limiting their rank.

\end{document}